# Strengthened Silicate Glasses by Residual Stress: Depth of Compression and Surface Flaws Stability Conditions.

Guglielmo Macrelli[1]

[1]*Isoclima SpA – R&D Department*

*Via A.Volta 14, 35042 Este (PD), Italy*

*guglielmomacrelli@hotmail.com*

[*]*Correspondence: guglielmomacrelli@hotmail.com*

**Abstract**

The application of silicate glasses in severe service environments requires a precise evaluation of structural strength under mechanical loads and surface tribological conditions. Because glass strength is governed by surface flaws and microcracks rather than being an intrinsic material property, residual surface compression fields, balanced by interior tensile zones, are widely implemented to inhibit flaw opening. Rather than relying on conventional allowable stress criteria to establish product acceptance, this study adopts a fracture mechanics framework based on the stress intensity factor $K_I$ and fundamental material limits: the critical stress intensity factor $K_{IC}$ for rapid fracture and the threshold stress intensity factor $K_{Ith}$ for time-delayed static fatigue failure. Using the Weight Function Method (WFM), $K_I$ is evaluated across generic surface flaw depths for two-dimensional continuous (2D-Continuous) surface cracks subjected to non-uniform internal residual stress fields and external loads. Flaw stability criteria are established for both zero-risk and moderate-risk design methodologies. Finally, the interaction and superposition of externally applied mechanical and thermal stresses with internal residual stress fields are evaluated.

## 1. Introduction

Silicate glass strength is not an inherent physical property, but it is strongly dependent on the presence of surface defects (surface flaws)[1]. If we consider strength values as indicated in Figure 1 we notice a huge dispersion over several orders of magnitude. Starting from the theoretical strength value that can be estimated from first principles[1], substantially related to the breakage of the silicon-oxygen chemical bond, we move from 35 GPa down to a few MPa for in-service highly damaged glass articles.

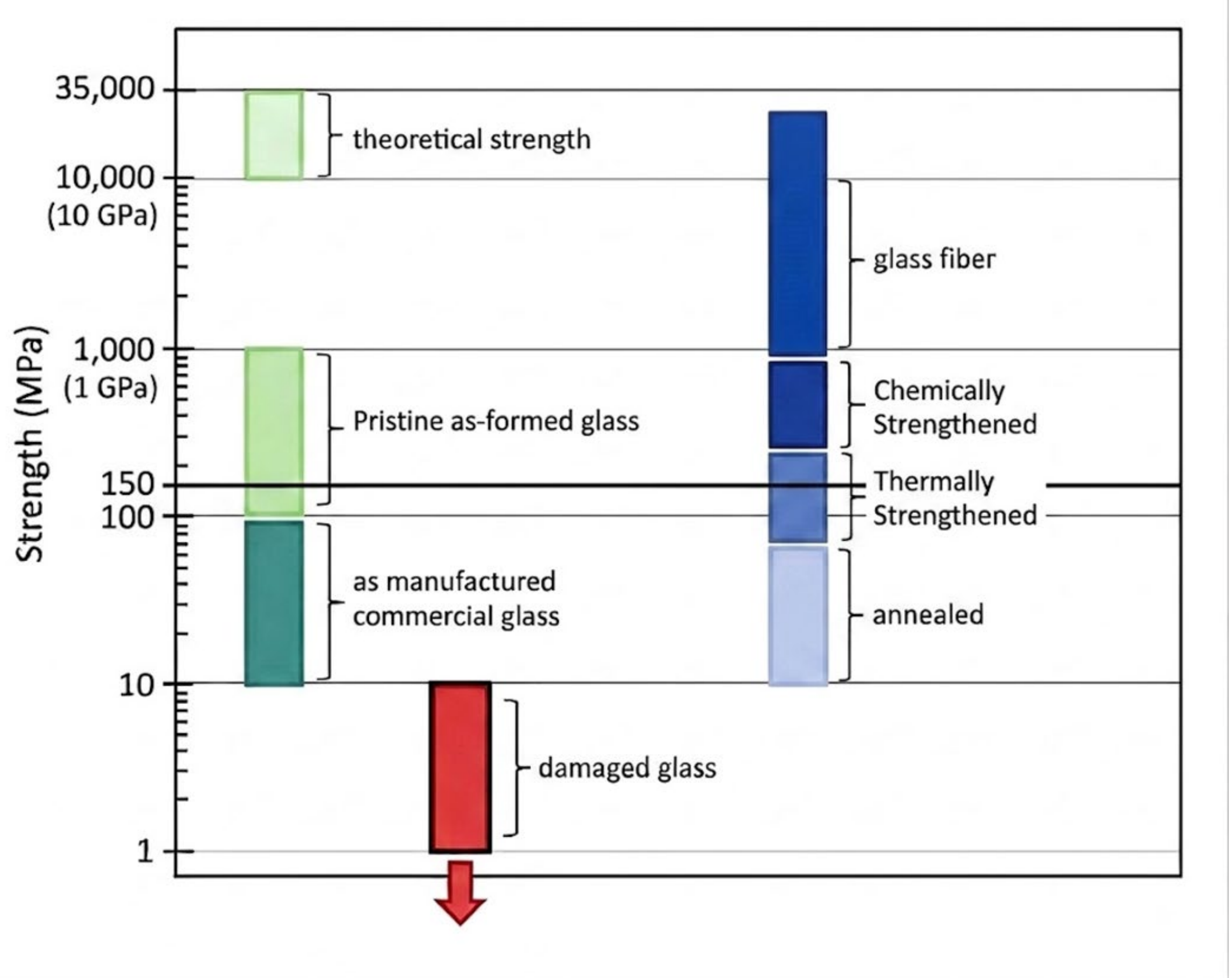

Figure 1 – Strength of silicate glass

On the right side of Figure 1, strength values for glass strengthened by thermal or chemical processes can be seen. In both strengthening methods, the strength increase concept is based on the introduction of a residual permanent stress field in the glass article's cross section. The idea is to put surface flaws under a compressive stress to prevent their opening when tensile external stress is applied. This concept is presented in Figure 2, where, in insert a), the amplification of applied stress through the surface flaw at the flaw tip is depicted, accounting for the huge strength reduction, while in insert b), the effect of a residual compressive stress is depicted working against the opening action of external tensile load. Different strengthening techniques were considered: the traditional thermal process based on the rapid quenching of the glass article from a uniform temperature above the glass transition value, and the chemical strengthening based on ion exchange, where a larger ion is introduced in the glass matrix, substituting an existing alkali, non-bridging ion. In glass breakage events, two failure conditions can be identified: immediate breakage and time-delayed fatigue breakage. In this study, failure conditions are evaluated in terms of fracture mechanics concepts: the critical stress intensity factor and the threshold stress intensity factor. The fracture mechanics concepts are introduced and integrated in a stress intensity factor versus crack depth diagram $K_I(a)$. The effect of introducing a residual stress is to shift the $K_I(a)$ curve in the negative part of the axis, and this is exactly the strengthening effect. The characteristics of the $K_I(a)$ curve identify different conditions of stability depending on the derivative sign of the curve. In the case of a negative derivative up to the minimum value, the more the crack increases its depth, the greater is the opposition of the compressive residual stress to the crack opening. This is a region of elevated stability. The other stability condition is up to the point where $K_I(a)$ becomes positive. A final limit stability condition is up to the onset of subcritical crack growth, that is, when $K_I(a)$ reaches the threshold value. It is recommended that, when relevant consequences may be generated by glass failure, the value of $K_I(a)$ is kept negative. The third condition of stability ($K_I(a)$ positive but below the $KIth$) is considered only when glass failure presents no consequences to the occupants, or it is not catastrophic in terms of damage. The advantage of this approach is to include at the same time both the stress resulting from externally applied mechanical or thermal loads and the residual stress resulting from technological strengthening processes. The nature of the different stresses is discussed and suitably modeled in mathematical formulations. The weight function method is finally used to evaluate the $K_I(a)$ diagram to be compared with the threshold limit and the final critical limit. In this study, the evaluation is limited to 2D edge cracks where the crack depth to glass plate thickness is negligible (semi-infinite plate approximation) and where the crack depth versus crack length is also negligible (edge crack). This approximation reduces the complexities in the calculation of the weight functions. Some further approximations in the calculation of the residual stress of ion-exchanged glass are presented and proved in a dedicated appendix. All calculations (residual stress profiles and stress intensity factor diagrams $K_I(a)$) are performed using PTC Mathcad Prime software (v. 10.0.0.0, Parametric Technology Corporation, Boston, MA, USA).

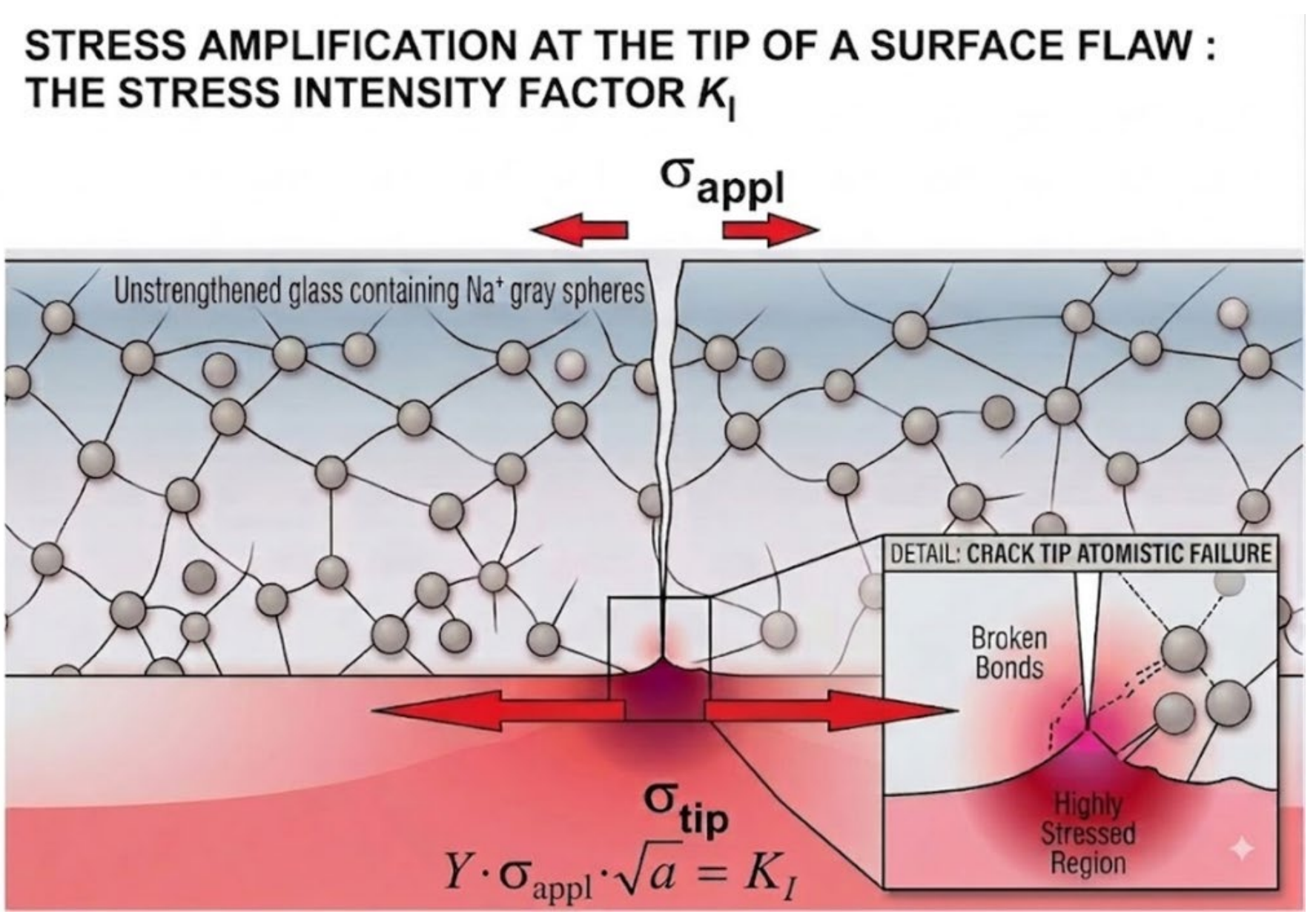

a)

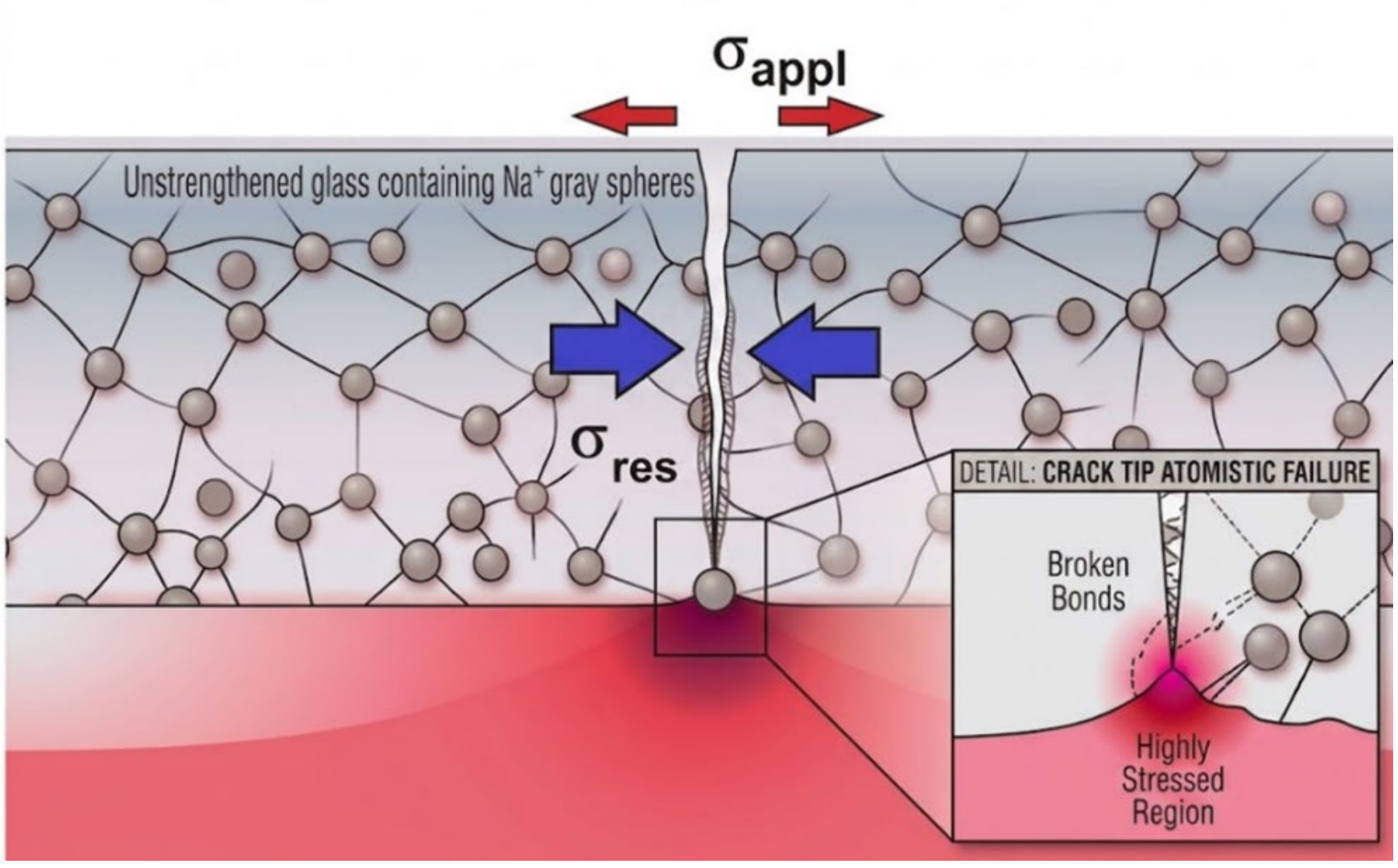

b)

Figure 2. a) Mechanism of amplification of applied stress at the surface flaw tip. b) Strengthening effect by the introduction of compressive residual stress

## 2. Fracture mechanics concepts. Silicate glass breakage conditions: rapid fracture and fatigue time-delayed fracture.

Figure 2 a) represents the main concept of how an applied load can be substantially amplified by a surface flaw up to the point where the amplified stress at the flaw tip reaches the theoretical strength of the material. Irwin[2,3,4,5] was the first to introduce the concept of stress intensity factor $K_I$ to connect the applied stress $\sigma_{appl}$ to some geometrical characteristics of the flaw: its depth $a$ and its shape effect represented by a non-dimensional parameter $Y$:

$$K_I = Y\sigma_{appl}\sqrt{a}\ . \qquad (1)$$

Equation (1) represents the case of an edge crack (depth $a$) in a semi-infinite plate under a uniform stress ($\sigma_{appl}$). As the applied stress increases, the stress intensity factor increases accordingly up to the point where the stress at the flaw tip reaches the value at which it breaks the silicon-oxygen bond, leading to immediate glass breakage. Figure 3 represents this event. The value of $K_I$ where we record immediate glass breakage is named the critical stress intensity factor $K_{IC}$, and the remarkable evidence is that it depends mostly on glass chemical composition. This means that we have finally identified a physical property that represents, at least on a macroscopic scale, an inherent strength characteristic of glass. The value of the applied stress corresponding to the critical stress intensity factor is the failure stress $\sigma_f$ of the glass. Equation (1) at breakage may be rewritten as the definition of $K_{IC}$:

$$K_{IC} = Y\sigma_f\sqrt{a^*}\ . \qquad (2)$$

In equation (2), a* is the value of the flaw depth at which breakage appears. This breakage mechanism is characteristic of brittle materials where, at the macroscopic level, no yield effects are recorded. The breakage event is clearly in correspondence with an external action represented by an increasing applied stress, and it develops on a pretty rapid time scale as applied stress reaches the $\sigma_f$ value.

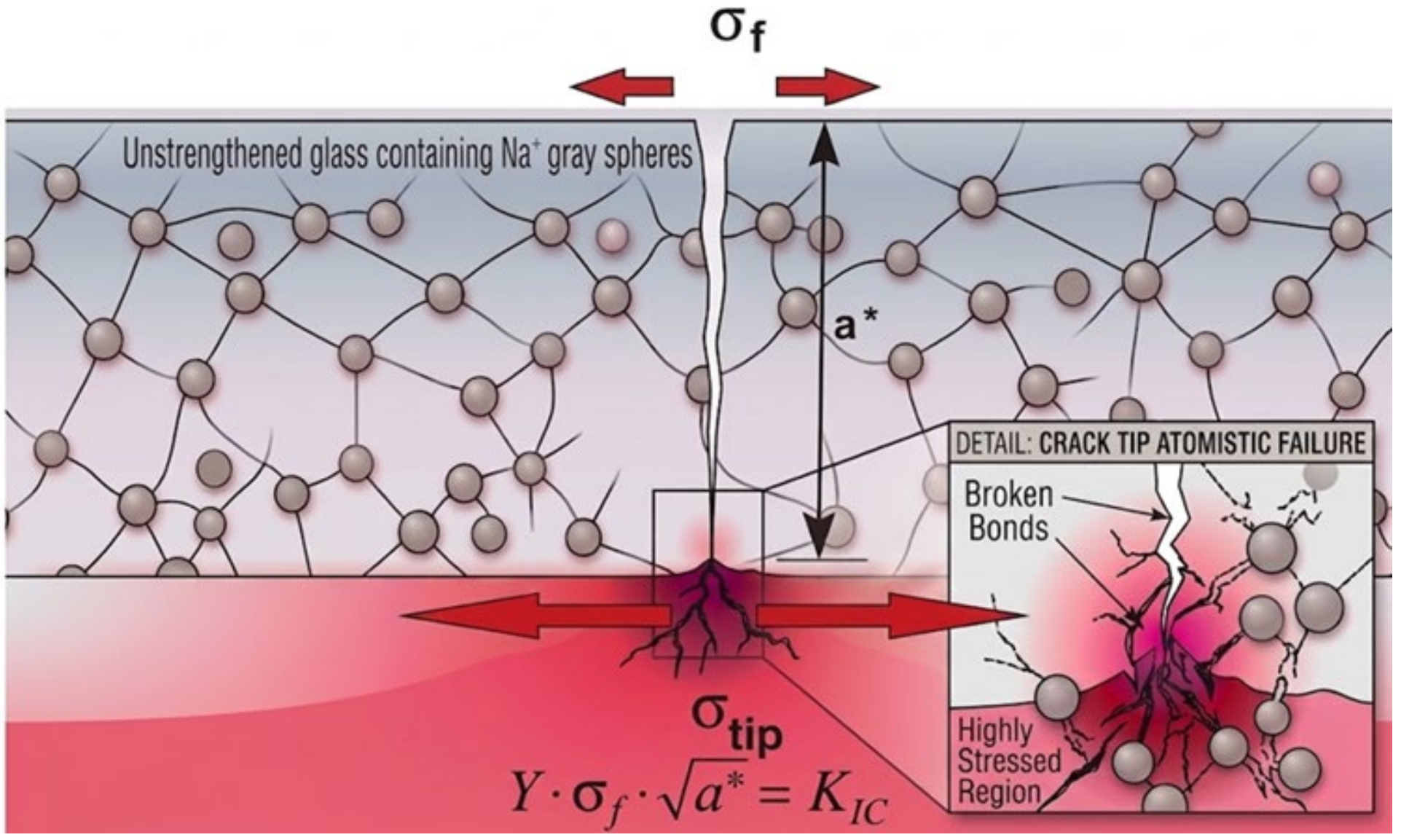

Figure 3 – Breakage event at the flaw tip characterized by the Critical Stress Intensity factor $K_{IC}$.

This is not the only fracture mechanism in silicate glasses. Static fatigue in silicate glasses is driven by subcritical crack growth (SCG), a process in which mechanical stress and chemical reaction with an aqueous environment act synergistically to break atomic bonds at the crack tip.

In the 1960s and 70s, Wiederhorn[6] formalized the quantitative study of this phenomenon by measuring crack velocity $v$ as a function of the stress intensity factor $K_I$, identifying three distinct regions in the $v(K_I)$ diagram (see Figure 4). In the Wiederhorn diagram, for a certain class of silicate glasses (alkali silicate glasses), a threshold value of $K_I$, named $K_{Ith}$, can be identified. When $K_I$ is below $K_{Ith}$ the crack velocity is zero. While $K_{IC}$ is an inherent characteristic of the glass, the $K_{Ith}$ value depends on environmental factors; for example, in liquid water or high humidity (>90% RH) $K_{Ith}$ drops by 10%-15% below the ambient air baseline, and in dry nitrogen or vacuum $K_{Ith} \rightarrow K_{IC}$ , that is, subcritical crack growth is suppressed. In 1982, Michalske and Freimann[7,8] integrated the macroscopic Wiederhorn model with an atomic-level physicochemical explanation for the rupture of strained siloxane (O-Si-O) bonds. A schematic diagram of the Michalske-Freimann model is depicted in Figure 5.

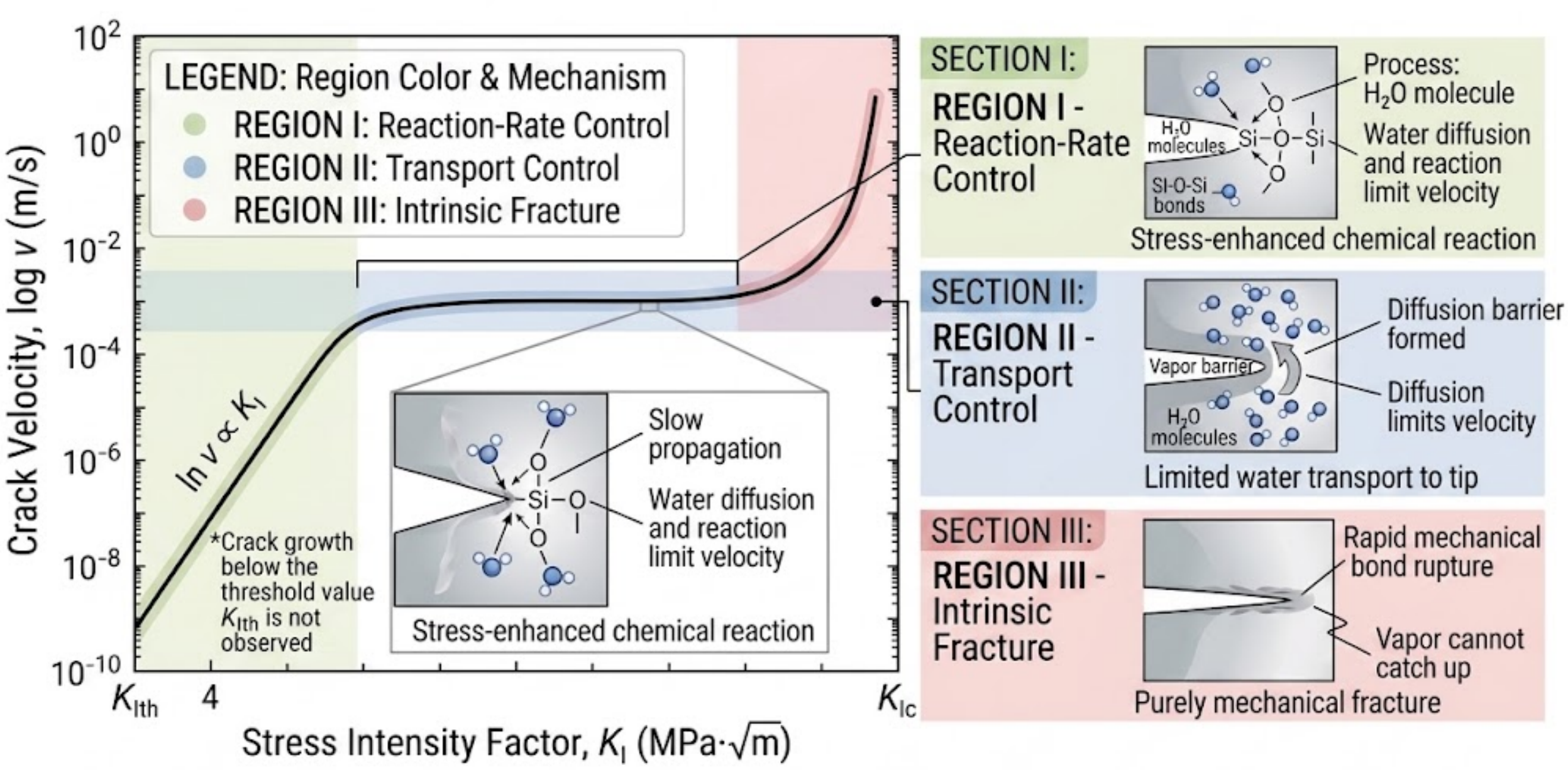

Figure 4 – Schematic Wiederhorn diagram for Subcritical Crack Growth in glass.

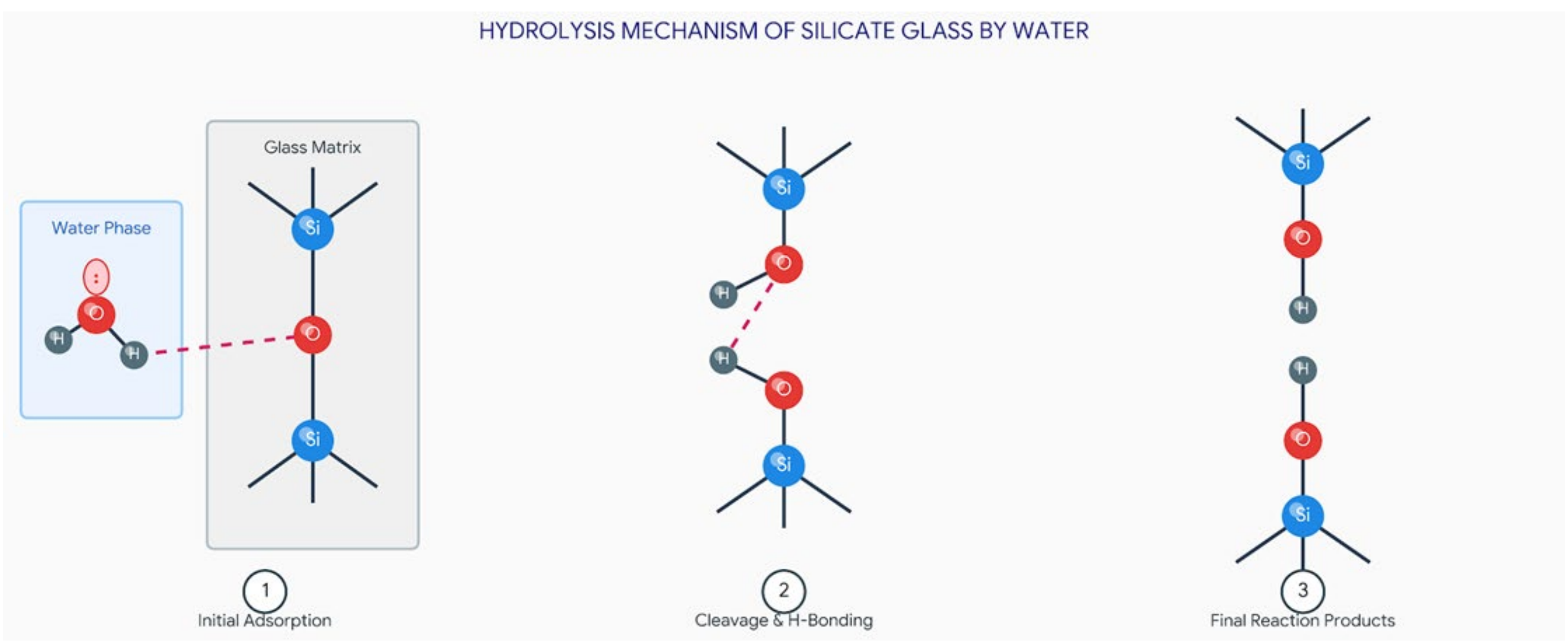

Figure 5 – The Michalske-Freimann atomic-level model

Coming back to the situation represented in Figure 2a) it can be outlined (see Figure 6) the effect of water-assisted cleavage leading to subcritical crack growth as $K_I$ reaches the $K_{Ith}$ value.

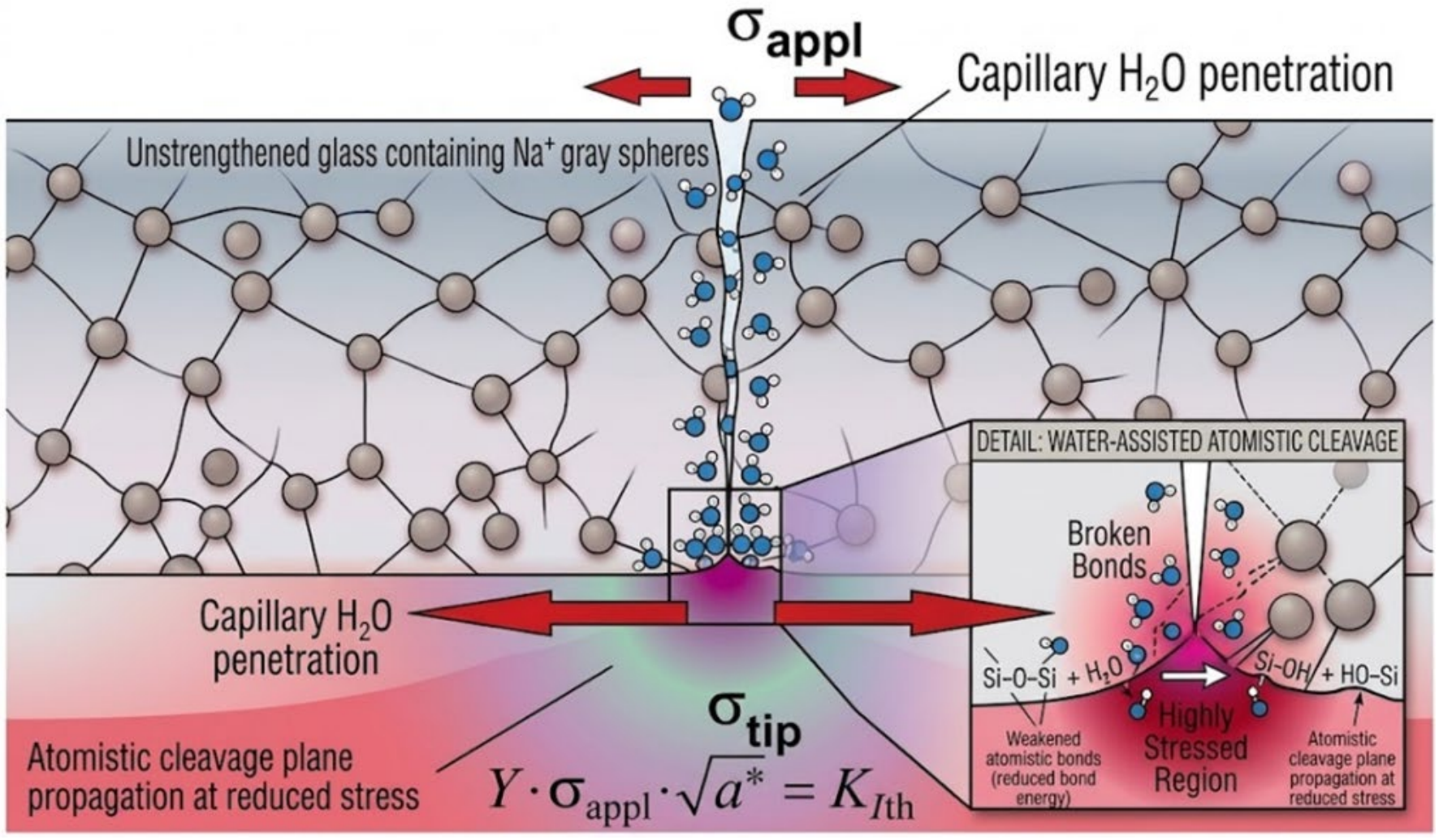

Figure 6 – Stress and water-assisted cleavage leading to SCG.

When silicate glass is subjected to tensile stress in a reactive environment like ambient moisture, cracks propagate at stress intensity levels significantly lower than the critical fracture one. The Wiederhorn diagram is a plot of crack velocity ($v=da/dt$) versus $K_I$ on a log-log scale. Distinct physical regimes can be identified:

A) When $K_I < K_{Ith}$ : the stress field at the crack tip energy is not sufficient to break Si-O-Si chemical bonds even in the presence of environmental moisture; the stress is not enough to lower the chemical activation energy required for water molecules ($H_2O$) to break the siloxane bonds. Crack velocity $v$ is zero, making the surface flaw mechanically stable indefinitely.

B) When $K_{Ith} \leq K_I << K_{IC}$ : This is a moisture reaction-rate-limited regime. Moisture diffuses to the crack and chemically assists bond cleavage at the crack tip, where stress is amplified and concentrated.

$$Si - O - Si + H_2O \rightarrow 2\left(Si - OH\right) \quad (3)$$

Applied tensile stress stretches and weakens atomic bonds, significantly lowering the activation barrier for hydrolysis. The crack propagates at subcritical speed following the Charles-Hillig[9] power law relationship:

$$v = \frac{da}{dt} = A\left(\frac{K_I}{K_{IC}}\right)^n, \tag{4}$$

where $A$ is a material/environmental constant and $n$ is the stress corrosion susceptibility index, which typically ranges $n \approx 15\text{-}30$ for silicate glass. Because $n$ is very large, crack speed is extremely sensitive to $K_I$. Small changes in surface stress or flaw depth, lead to order-of-magnitude changes in crack velocity.

C) When $K_I$ overpasses the reaction rate-limited regime but does not reach the critical value, a transport/diffusion limited plateau is observed. The plateau is originated by a Transport/Diffusion limited physical mechanism: the crack grows so fast that environmental water molecules cannot diffuse to the crack tip fast enough to keep pace with bond cleavage. As a result, crack velocity is nearly constant ($v \approx const$), practically independent of $K_I$ and entirely governed by ambient humidity, solvent concentration, and molecular transport rates.

D) When $K_I \geq K_{IC}$ : In this regime, mechanical stress alone is sufficient to cleave atomic bonds and environmental chemical assistance is no more needed to cleave atomic bonds. Crack propagation becomes unstable and approaches terminal velocity ($v \approx 1500\text{-}2000\ m/s$), this is the fast-fracture regime.

It is interesting to point out some consequences of the characteristic crack-velocity power law (equation (4)). If we indicate with $a$ the initial flaw depth and by $a^*$ the critical flaw depth at breakage, we can define a time-to-failure expression:

$$t_{fail} = \int_a^{a^*} \frac{1}{v(K_I(a))} da = \int_a^{a^*} \frac{1}{A\left(\frac{K_I(a)}{K_{IC}}\right)^n} da \ . \tag{5}$$

Due to the steep n-power dependence of $v$ with $(K_I)^n$, crack velocity at $a$ is drastically smaller than near $a$*. Consequently, more than 90% of the total glass article lifetime is spent growing the crack through the first few nanometers/micrometers near $a$. Once the crack exceeds $a$, the remaining time to catastrophic failure is negligible. A second consequence is that, under a uniform stress $\sigma$, where equation (1) is valid, integrating equation (5) leads to $t_{fail} \rightarrow (\sigma)^{-n}$ . Because $n$ is typically around 20 for silicate glasses, an increase of 10% in stress reduces time to failure by a factor 6.7, cutting service life by 85%. Doubling applied stress reduces service life by one million ($2^{20} \approx 10^6$). The conclusion of the above discussion is that to evaluate glass breakage, we have to consider two characteristics: the $K_{IC}$ and the $K_{Ith}$ values.

### 3. Characteristics of externally applied loads.

As already discussed, in this study, an externally applied load will be considered, generating stress $\sigma_{appl}(x)$ in a glass plate. Service life actions on glass articles can be either mechanical or thermal. The simplest case is a uniform stress, as discussed when we introduced equation (1). The other case of mechanical stress on a flat glass plate is pure bending. In this last case, normal stress varies linearly across the plate thickness $d$ from maximum tensile value $(+SM)$ at one surface to maximum compression $(-SM)$ at the opposing surface, crossing zero at the mid-plane. This forms a linear antisymmetric profile through the plate thickness. The peak surface bending stress is given by:

$$SM = \frac{6M}{d^2}, \qquad (6)$$

where $M$ is the applied bending moment per unit width, and $t$ is the plate thickness. The applied stress distribution as a function of the plate coordinate $x = [0,d]$ is:

$$\sigma_{appl}(x) = SM\left(1 - 2\frac{x}{d}\right). \qquad (7)$$

Thermal stress can be either due to a through-thickness thermal gradient or to an in-plane thermal gradient. In the first case, stress arises only when the glass plate is constrained; otherwise, the unconstrained condition accommodates a linear gradient through curvature, resulting in zero internal stress. For a plate constrained against bowing with a linear temperature difference ΔT, the maximum surface stress magnitude is:

$$SM = \frac{E\alpha\Delta T}{2(1-\nu)}, \tag{8}$$

where $E$ is the Young's modulus, $\nu$ the Poisson's ratio, $\alpha$ the linear thermal expansion coefficient. In the case of an in-plane thermal gradient, temperature variations occur across the glass plane. We may assume that at any specific surface point, the temperature is uniform through the thickness $x$. As a consequence, the stress profile through the thickness $x$ may be considered uniform across the flaw depth $a$. The peak tensile edge stress driven by in-plane temperature difference $\Delta T = T_{center} - T_{edge}$ is:

$$\sigma_{InpLane} = f \cdot \frac{E\alpha\Delta T}{(1-\nu)}, \tag{9}$$

where $f$ is a structural constraint factor ($0.5 \leq f \leq 1.0$) determined by edge-frame insulation, aspect ratio, and plane geometry. In this study, we consider an applied stress as uniform over the plate thickness or linearly variable according to equation (7).

## 4. Silicate Glass physical characteristics and related chemical structural mechanisms

In the above discussion, several physical characteristics of silicate glasses ($K_{IC}$, $K_{Ith}$, $n$, $E$, $\nu$, $\alpha$) have been introduced. These characteristics govern fracture mechanics, subcritical crack growth, and stress generation under externally applied loads. Herewith the discussion is limited to the main families of silicate glasses generated from alkali silicates: Soda-lime Silicate (SLS), Sodium Borosilicate (SBS), Sodium Aluminosilicate (SAS), Sodium Aluminum Borosilicate (SABS), Lithium Aluminosilicate (LAS), Lithium Alumino Borosilicate (LABS). The reason for considering these glasses is that they are the main glass types used for chemical strengthening by ion exchange and because, in most applications, thermal strengthening by rapid thermal quenching is applied to SLS and SBS glasses.

We conclude that the main technological processes (Ion Exchange and Thermal Quenching) used to introduce residual stress in glass articles to increase strength are applied to Alkali Silicate glasses. In Table I, the values of the physical characteristics of silicate glasses are reported.

Table I – Physical characteristics of main families of Alkali Silicate glasses

| Glass Family | Acronym | $K_{IC}$(MPa√m) | $K_{ITh}$(MPa√m) | n | E(GPa) | Poisson Rato –ν | α(10-6/K) |
|---|---|---|---|---|---|---|---|
| Soda-Lime | SLS | 0.70-0.78 | 0.25-0.32 | 16-20 | 68-74 | 0.21-0.23 | 8.5-9.5 |
| Sodium Borosilicate | SBS | 0.75-0.82 | 0.30-0.38 | 25-35 | 60-68 | 0.19-0.21 | 3.0-5.5 |
| Sodium Aluminosilicate | SAS | 0.75-0.85 | 0.35-0.45 | 20-28 | 68-76 | 0.21-0.23 | 7.5-8.8 |
| Sodium AluminoBorosilicate | SABS | 0.78-0.88 | 0.35-0.45 | 22-30 | 65-73 | 0.20-0.22 | 5.0-7.0 |
| LithiumAluminosilicate | LAS | 0.85-1.05 | 0.40-0.50 | 22-32 | 80-90 | 0.21-0.24 | 4.0-6.0 |
| Lithium AluminoBorosilicate | LABS | 0.82-0.95 | 0.38-0.48 | 24-32 | 76-84 | 0.22-0.26 | 4.0-5.5 |

Notes to Table I:
(*) Values are indicated with ranges that depend on actual chemical composition (mainly alkali concentration)
(**) Subcritical threshold values (KIth) refer to ambient air (40%-60% RH at 20°C-25°C)
(***) In case of Strengthening by Ion Exchange, all properties are referred to the un-exchanged parent glass matrix.

Looking to Table I it can be found an evolution of alkali silicate glass chemical compositions can be found, where the chemical and structural changes generate effects that influence fracture and mechanical properties. In most cases (SAS and LAS), the choice was driven by other motivations (for example, in Ion Exchange strengthening processes, a significantly higher interdiffusion coefficient of parent alkali versus incoming alkali); nevertheless, it is remarkable that other relevant effects were generated, playing a significant role in mechanical fracture behavior. In Table II, some remarkable effects resulting from the chemical structural mechanisms are highlighted:  alumina incorporation, Lithium field strength, and Boron anomalies.

Table II – Effects on physical characteristics resulting from Chemical Structural changes

| Chemical Structural Feature | Effect |
|---|---|
| Alumina ($Al_2O_3$) Incorporation | Converts Non-Bridging Oxygens into fully cross-linked [AlO4]- tetrahedral networks. This hinders moisture transport at crack tip, elevating KIth from 0.25 in SLS to 0.38 MPa√m in SAS. |
| Lithium (Li+) Field Strength | Lithium has a smaller ionic radius (0.76 A) compared to Na+ (1.02 A) providing in this way a higher field strength and atomic packing density. This raises Young Modulus (E≈85 GPa) and intrinsically boosts fracture toughness (KIC ≈0.90 MPa√m) |
| Boron ($B_2O3$) Anomalies | Trigonal [BO3] to tetrahedral [BO4] coordination increases structural compliance under hydrostatic and shear stress, raising the stress corrosion susceptibility index n > 25. Higher n steepen the v(KI) slopes resulting in higher fatigue resistance near KIth. |

:

## 5. Characteristics of residual stress profiles

It has already been pointed out that the widely used approach to increase glass strength is the introduction of a residual stress in the glass article cross section with the characteristic of having a compression stress state in the near surface area balanced by an internal tensile state. Figure 2a) represents the concept, while Figure 7 indicates a clearer picture where the residual stress is a non-uniform stress field that keeps the surface flaws under a compression state, preventing their opening under an externally applied stress. The widely used methods to introduce residual stress in a glass article are thermal strengthening where, by rapid quenching from a temperature above the glass transition temperature, a roughly parabolic residual stress is introduced [1,10,11] and chemical strengthening where, by ion-exchange[1,11,12,13], a larger network modifying ion (usually an alkali ion) is exchanged for a smaller network modifying ion (again an alkali), resulting in a non-uniform residual stress characterized by a surface compression (value of stress field at glass surface) and a compression layer depth (*Cd* or *DOL*) which is the distance from glass surface to the point where residual stress changes from compression to tensile *($\sigma(Cd)=0$)*.

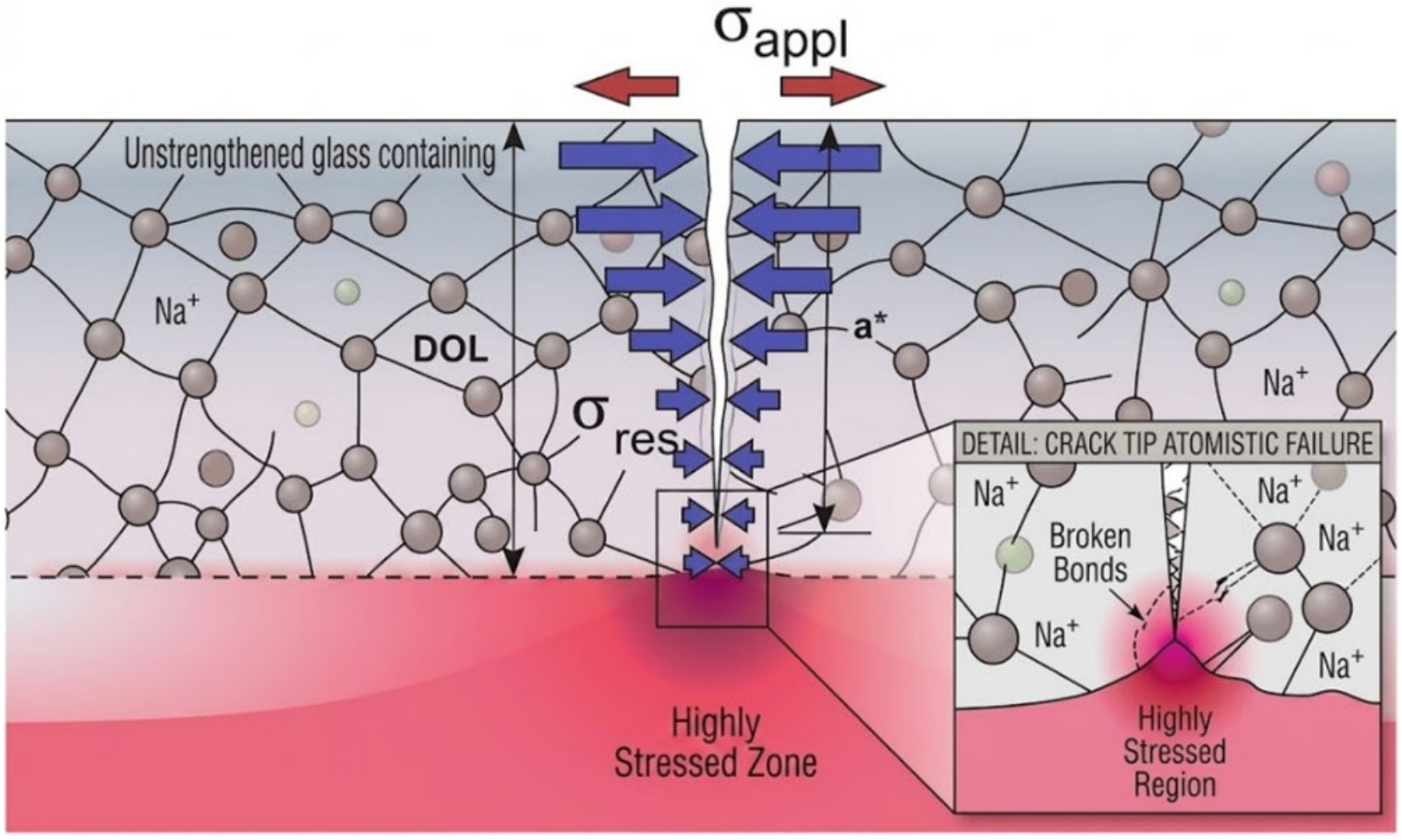

Figure 7 – Non-Uniform Residual stress field (blue arrows) generating a compression state along a surface flaw subjected to a tensile stress generated by an external load.

In this study they will be considered both residual stresses coming from thermal strengthening processes (fast quenching of glass articles from temperatures above the glass transition *Tg*) and ion exchange processes performed below glass transition temperatures, where a large alkali ion is exchanged with a smaller non-bridging alkali ion in the glass matrix. The residual stress resulting from thermal processes (TSG) is modeled as a parabolic curve[11]:

$$\sigma_{res}^{TSG}(x) = 12 \cdot CT^{TSG} \cdot \left( \frac{x}{d} - \frac{x^2}{d^2} - \frac{1}{6} \right), \tag{10}$$

where $CT^{res}$ is the value of central tension at mid-glass thickness $d/2$. At $x=0$ and $x=d$ (glass article surfaces) it results:

$$\sigma_{res}^{TSG}(x=0, x=d) = -2 \cdot CT^{TSG}. \tag{11}$$

The values of x where the residual stress is zero are the solutions of the second-order algebraic equation (10):

$$\sigma_{res}^{TSG}(x) = 0, \tag{12}$$

$$x_1 = 0.2113 \cdot d \,;\; x_1 = 0.7887 \cdot d\,. \tag{13}$$

If we consider a glass article of thickness *d= 4 mm* and a central tension $CT^{TSG}$*=45 MPa,* the residual stress plot according to equation (10) is reported in Figure 8.:

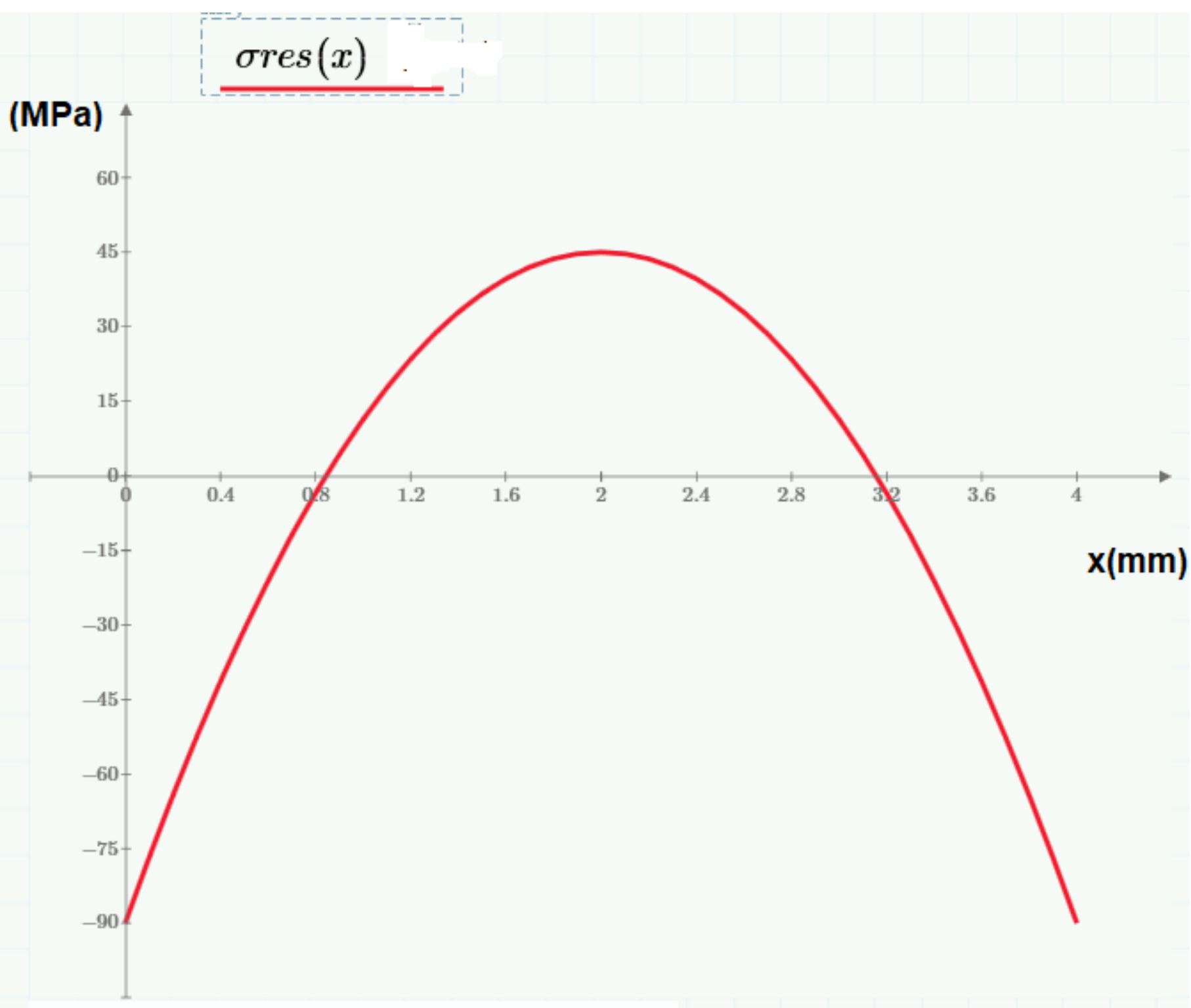

**Fi**gure 8 – Residual stress of a 4mm glass article with a surface compression *SC*= 90 MPa and a central tension of 45 MPa according to equation (10). The compression layer is *Cd*=0.845mm.

The residual stress resulting from ion exchange can be modeled in a more complex way[13,14] starting from the residual concentration $c(x,t)$ $(mol/cm^3)$ of the incoming ions in the glass:

$$c(x,t) = C_s erfc\left(\frac{x}{2\sqrt{D \cdot t}}\right), \qquad (14)$$

where *Cs* is the value of the surface concentration of the incoming ions (considered constant, because it is the boundary condition for this particular solution), *t* is the total time of the ion-exchange interdiffusion process, and *D* is the interdiffusion coefficient. Residual stress, as a consequence of ion exchange, is built up according to the mechanical scheme reported in Figure 9a.

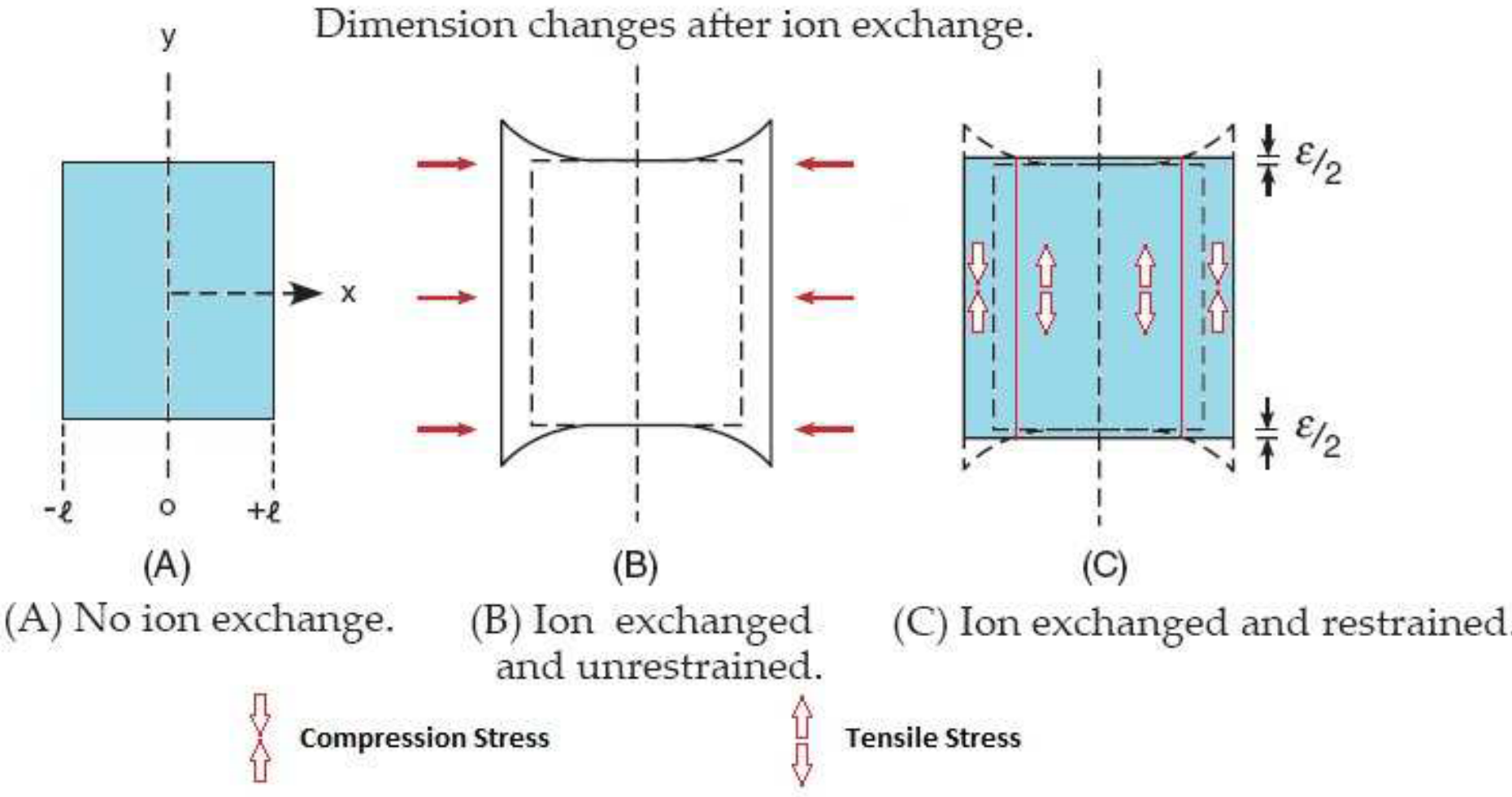

a)

b)

Figure 9 – Stress build-up by ion exchange in glass articles: a) principle of stress formation by the enforcement of compatibility criteria b) resulting equi-biaxial stress.

The larger incoming ions determine a near-surface molar volume expansion of the glass $\Delta Vm$. This volumetric expansion is constrained by the inner part of the glass not exposed to ion exchange. The enforcement of compatibility criteria (Figure 9a) insert (C)) determines a near-surface compression balanced by an inner tensile state. The nature of this stress build-up mechanism results in a typical equi-biaxial stress: $\sigma_{xx}=0$, $\sigma_{EB}=\sigma_{yy}=\sigma_{zz}$ as depicted in Figure 9b. The mathematical model of stress build-up as a consequence of ion exchange follows the seminal paper of Sane and Cooper[15] and can be expressed through equation (15a) that was first derived by Macrelli[16] from the Sane Cooper original equation[15] using the integration by part formula and further used by Macrelli, Varshneya and Mauro[17] for the resolution of the Donald and Hill anomaly and by Macrelli and others[13] to model stress profile of Soda Lime and Sodium Aluminosilicate glass.

$$\sigma_{EB}(x,t) = -\frac{B \cdot E \cdot V}{1-\nu}\left[\left(c(x,t)-\bar{c}(t)\right)-\int_0^t \frac{\partial R(t-\theta)}{\partial \theta}\left(c(x,\theta)-\bar{c}(\theta)\right)d\theta\right], \quad (15a)$$

$$R(t) = \exp\left[-\left(\frac{t}{\tau}\right)^b\right] \quad . \quad (15b)$$

In equation (15a) $E$ is the Young modulus, $n$ the Poisson ratio, $B$ is the network expansion coefficient[15] and $V$ is the Varshneya factor[16,17,18,19]. The Varshneya factor $V$ is justified[16,17] because the incorporation of larger incoming ions (Potassium, for example) into the silicate network (in place of smaller alkali like Sodium) induces a near-instantaneous local rearrangement (on nanosecond to picosecond timescales) that both the unrelaxed elastic modulus $E$ and classical $\alpha$-viscoelasticity fail to capture. The Varshneya factor is non-dimensional and presents values ranging from 0.6 to 0.65[13] for SLS and SAS glasses. Multiplying by $V$, the front term of the stress equation, effectively scales down the theoretical magnitude of the stuffing effect to a level that the glass network can realistically sustain. The function $R(t)$ is the relaxation function accounting for α-viscoelastic relaxation effects, and it is usually expressed in the KWW stretched Maxwellian form[1,13,17]. In (15b), $\tau$ is the relaxation time of the viscoelastic relaxation process and b is the exponent of the stretched function. The relaxation function is a limited function of time : $0 \leq R(t) \leq 1$ and $R(0)=1$ (No relaxation at initial time t=0), while we have full relaxation at infinite time ($R(t\rightarrow\infty)=0$). Considering a 4 mm-thick glass article, we can plot the residual stress for two different glass types (SLS and SAS). The resulting stress profile is depicted in Figure 10 where the stress profile is calculated according to equation (15a) and profile characteristics are reported in Table III.

Table III – Residual stress profile characteristics of SLS and SAS silicate glasses.

| Glass Type | Process data: Time /Temperature | Surface Compression Sc(MPA) | Case Depth Cd(μm) |
|---|---|---|---|
| SLS | 24h/450°C | 463 | 35 |
| SAS | 72h/450°C | 727 | 290 |

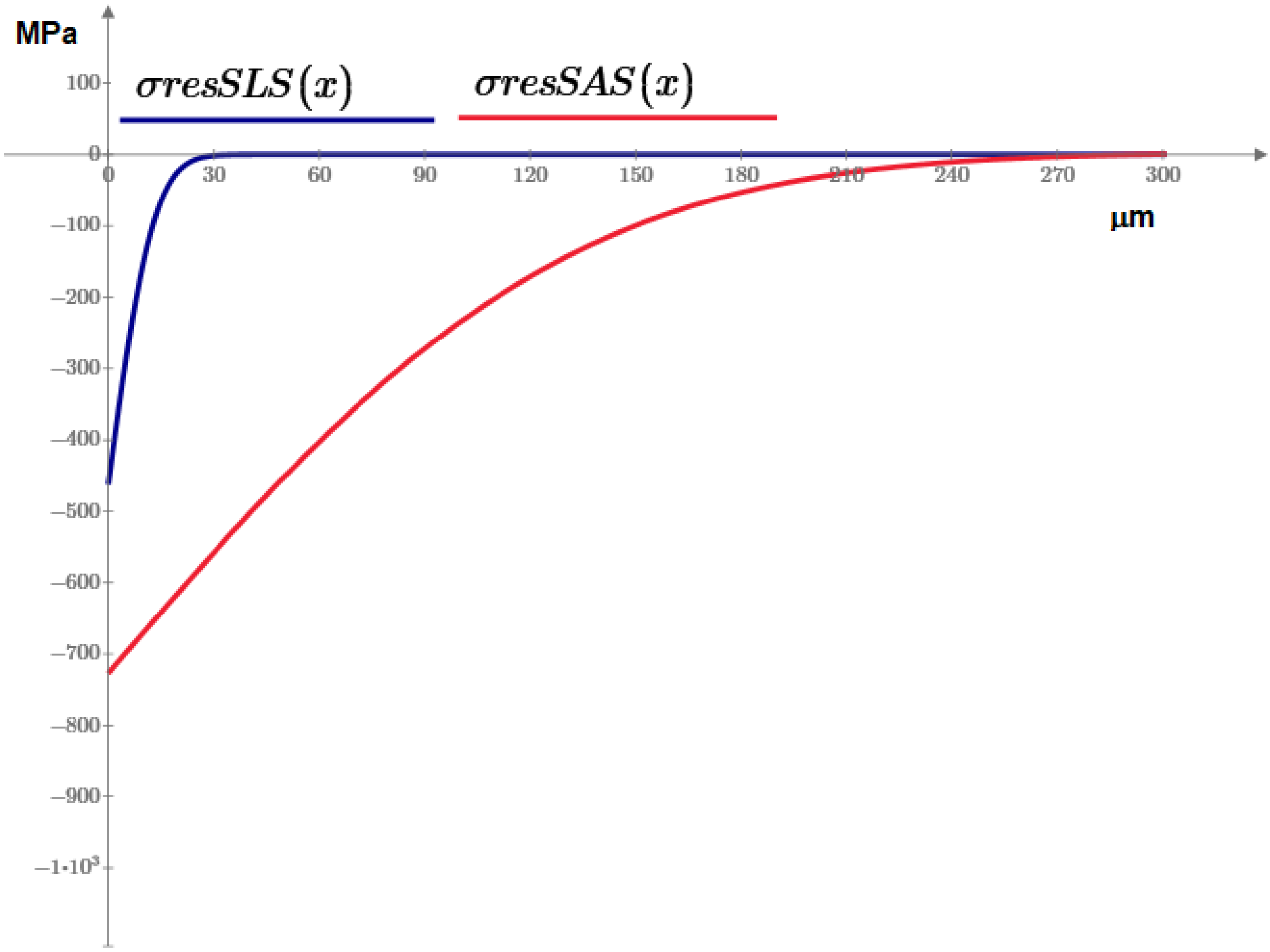

Figure 10- Residual stress of a 4mm glass article of SLS and SAS types according to the characteristics of Table III.

The linear network dilatation coefficient (LNDC) B is defined as the relative molar volumetric expansion per unit concentration of exchanged ions:

$$B = \frac{\Delta V_m}{3V_m C}. \qquad (16)$$

And has units of ($cm^3$/mol). Defining the surface compression developed at zero time:

$$ScV = \frac{B \cdot E \cdot V \cdot c(0,t)}{1-\nu} = \frac{B \cdot E \cdot V \cdot C_s}{1-\nu}. \tag{17}$$

Using equation (15a), it can be proved, under a wide approximation condition, that the relaxation effect allows evaluating the time evolution of surface compression:

$$Sc(t) = ScV \cdot R(t). \tag{18}$$

The result expressed in equation (18) suggests a further approximation of equation (15a). When the relaxation time of the relaxation function is larger than the observation time, in other words, if relaxation can be considered extremely slow (τ<<t), the convolution integral term in equation (15a) can be decoupled (*R(t-θ)≈R(t)*), resulting in:

$$\sigma_{EB}(x,t) \approx -\frac{ScV}{C_s} \cdot R(t)\left[\left(c(x,t) - \bar{c}(t)\right)\right] = -\frac{Sc(t)}{C_s}\left[\left(c(x,t) - \bar{c}(t)\right)\right] \tag{19}$$

In Appendix 1, we derive the two approximations in equations (18) and (19) from equation (15a). Where stress relaxation effects are not critical, that is, for SLS and SAS glasses[13], the approximation expressed by equation (19) provides an acceptable estimation of residual stress profiles generated by ion exchange.

### 6. Stress intensity factor calculation by weight functions method (WFM)

The crucial point in glass strength evaluation is calculating the stress intensity factor $K_I$ as a function of surface flaw depth $a$. The strength assessment is then performed by comparing $K_I(a)$ with the critical value $K_{IC}$ and the threshold value $K_{Ith}$, which trigger subcritical crack growth (SCG). The calculation of the stress intensity factor under a non-uniform stress field σ(x) was greatly simplified in the 1970s of the last century after a seminal paper by Bueckner[18,] and a breakthrough theorem that was demonstrated by Rice[19].

The mechanics of fracture in materials with non-uniform stress distributions, such as thermally tempered or chemically strengthened glass, rely directly on the weight function method established by Hans Bueckner and James R. Rice in the early 1970s. Before the 1970s, calculating the Mode-I stress intensity factor ($K_I$) for a crack embedded in an arbitrary, non-uniform stress field $\sigma(x)$ required solving complex, geometry-specific elastic boundary-value problems from scratch for every unique stress profile. Hans Bueckner[18] demonstrated that for a given crack geometry, there exists a unique, universal weight function $m(x,a)$ that depends solely on geometry and boundary conditions, completely independent of the applied loading. Using Green's function approach, the stress intensity factor under any arbitrary stress profile $\sigma(x)$ along the crack line reduces to a straightforward integral:

$$K_I\left(a\right) = \int_0^a \sigma(x) \cdot m(x,a)dx \,. \tag{19}$$

James R. Rice[19] provided the key breakthrough that made Bueckner's weight functions practical. Rice proved that the weight function $m(x,a)$ can be obtained directly by differentiating the crack surface displacement field $u_r(x,a)$ of any single, simple reference loading case with respect to crack length $a$:

$$m(x,a) = \frac{E'}{2K_r}\frac{\partial u_r(x,a)}{\partial a}, \tag{20}$$

where $E'$ is the effective elastic modulus and $K_r$ is the stress intensity factor of the reference loading state. While the Bueckner-Rice approach represented by equations (19) and (20) was conceptually a breakthrough, it still presented some practical limitations because it required knowing a reference displacement field. This limitation was eliminated by another crucial paper by Petroski and Achenbach[20] where they proposed an approximate representation for the crack opening displacement profile $u_r(x,a)$ that requires only the reference stress intensity factor $K_r$. The significance and impact of the Petroski-Achenbach contribution is in the elimination of the requirement of the full displacement field, making weight functions far more accessible for computing $K_I$ under complex, non-uniform stress profiles. This short historical description of the genesis of the weight function method (WFM) indicates the theoretical foundation of the subsequent multi-parameter weight function approximations such as those developed by Tada-Paris-Irwin[5], Shen- Glinka[21,22] , Wu-Carlssson[23], Wang-Lambert[24]and Fett and Munz[25]. After the seminal papers of Bueckner[18], Rice[19], and Petroski and Achenback[20] the problem has been extended to the calculation of the $K_I$ for a three-dimensional semi-elliptical surface crack as represented in Figure 11 a) and b). In the modern developments of the weight functions method for the calculation of the $K_I$ for a three-dimensional semi-elliptical surface crack, two approaches can be identified. The first one finds its theoretical foundation in Bueckner-Tada, where a two-parameter ($M_1$ and $M_2$) integer-exponent function is used:

$$m(x,a)=\frac{2}{\sqrt{2\pi(a-x)}}\left[1+M_1\left(1-\frac{x}{a}\right)+M_2\left(1-\frac{x}{a}\right)^2\right]. \tag{21}$$

This formulation, originally proposed for a 2D case, has been further extended by Wu-Carlsson[23] to the 3D case by formulating $M_1$ and $M_2$ based on crack surface displacement fields $u(x,a)$ and reference solutions. $M_1$ and $M_2$ are expressed as polynomial surfaces fitted against reference $K_I$ datasets. The other approach has been proposed by Shen and Glinka[21],[22] based on a three parameters ($M_1$, $M_2$ and $M_3$) fractional-exponent functions:

$$m(x,a)=\frac{2}{\sqrt{2\pi(a-x)}}\left[1+M_1\left(1-\frac{x}{a}\right)^{1/2}+M_2\left(1-\frac{x}{a}\right)^{1}+M_3\left(1-\frac{x}{a}\right)^{3/2}\right] . \tag{22}$$

The Shen-Glinka approach has been further validated by Wang and Lambert where it has finally assessed how the $Mi$ coefficients are determined by a direct comparison with reference $K_I$ calculation (uniform tensile and pure bending) and using the Fett[25] open-mouth crack criteria. In Tables IV and V the historical developments, from the early breakthrough of the 1970s (Table IV) to the modern assessment of the 1990s (Table V), have been summarized. In this study, in order to reduce complexities, we consider a 2D edge crack, which means considering the aspect ratio *a/d* and crack depth/thickness a/c to zero (see Figure 11). The advantage is that the *Mi* coefficients become fixed numbers. Both approaches: integer functions (Equation 21) and fractional functions (Equation 22) work for the 2D limit, providing results for $K_I$(a) which deviates less than 1%; hence, they can be considered equivalent. The weight function method (WFM) is used to determine the stress intensity factor $K_I(a)$ as a function of the surface crack depth (*a*) for 2D edge crack with a total non-uniform stress generated by external loads $\sigma_{appl}(x)$ and internal residual stress $\sigma_{res}(x)$ generated by thermal and ion exchange processes. The total stress field is:

$$\sigma_T(x) = \sigma_{appl}(x) + \sigma_{res}(x), \tag{23}$$

and the stress intensity factor:

$$K_I(a) = \int_0^a \sigma_T(x) \cdot m(x,a)dx, \tag{24}$$

where the *m(x,a)* weight function can be either the Tada-Wu_Carlsson[21] or the Shen-Glinka-Wang-Lambert[22] function with *Mi* coefficients for the 2D Edge crack.

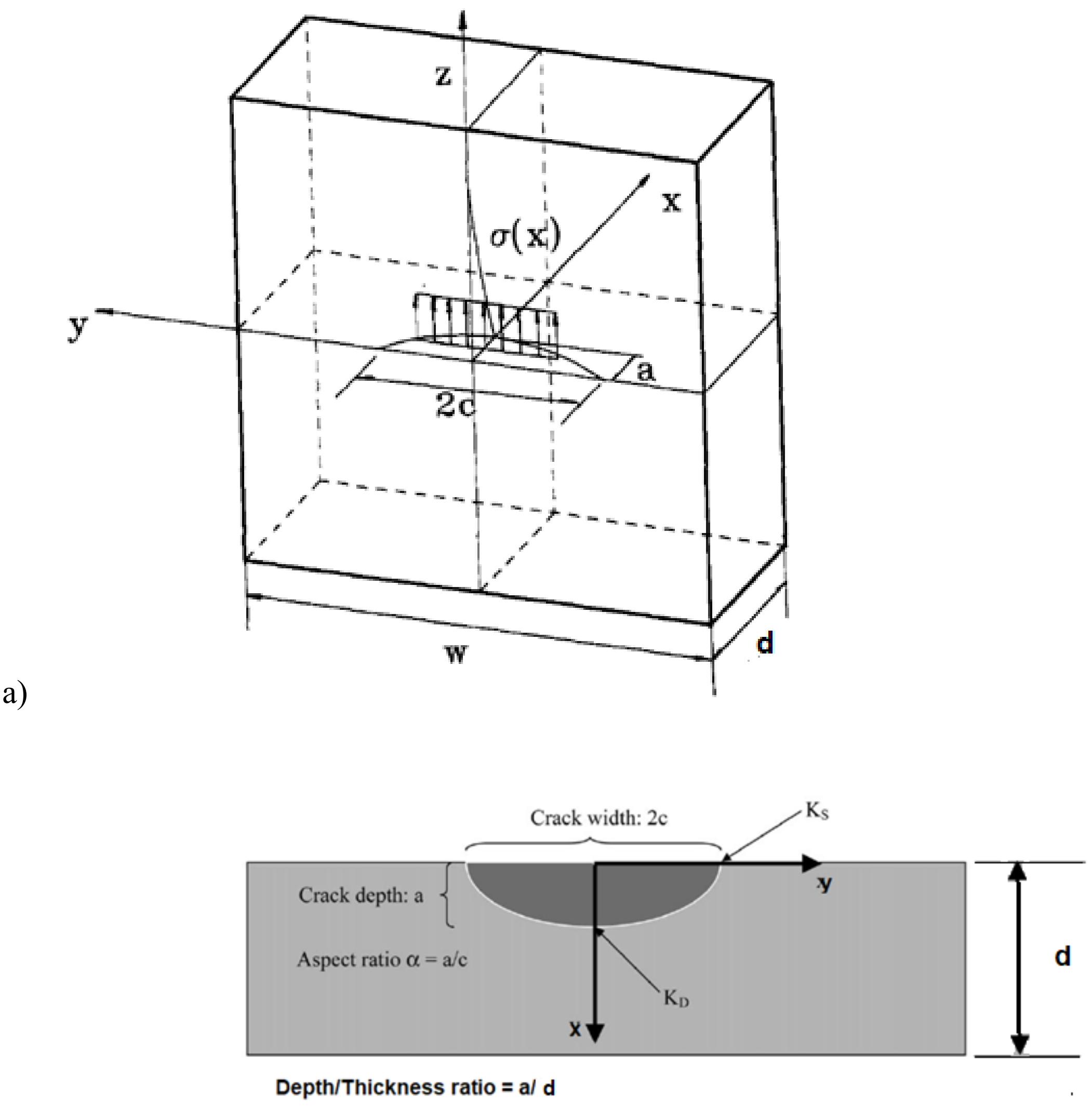

b)

Figure 11 – 3D Semi-elliptical surface crack – a) 3D representation and coordinate system b) projection on the *x,y* plane.

**Table IV. Historical evolution and comparison of weight function formulations in Linear Elastic Fracture Mechanics (LEFM) – Late developments 1970's**

| Milestone / Authors | Year | Mathematical Basis | Required Inputs / Calibration | Key Scientific Contribution |
|---|---|---|---|---|
| **Bueckner**[18] | 1970 | Crack-face displacement field derivative: $m(x,a) \propto \frac{\partial u_r(x,a)}{\partial a}$ | Exact analytical displacement solution ur(x,a) for a reference loading | Established that the weight function is an intrinsic geometric property, independent of applied loading. |
| **Rice**[19] | 1972 | Energy release rate (G) / J-integral formulation: $m(x,a) \propto \frac{\partial u_r(x,a)}{\partial A}$ | Reference $K_I$ and displacement variation over crack area increment ∂A | Generalized weight function theory from 2D planar geometries to arbitrary 3D crack fronts. |
| **Tada, Paris & Irwin**[5] | 1973 | 2D integer power series expansion: $\left(1-x/a\right)^1, \left(1-x/a\right)^2, \ldots$ | Tabulated 2D reference stress intensity factors ($K_I$) | Standardized analytical 2D weight function expressions for engineering handbooks. |
| **Petroski & Achenbach**[20] | 1977 | Parametric Crack Opening Displacement (COD) profile | Single reference $K_I$ + Global energy conservation balance | Eliminated the need for exact displacement fields, enabling practical analytical weight function derivation. |

**Table V. Historical evolution and comparison of weight function formulations in Linear Elastic Fracture Mechanics (LEFM) – Modern developments 1990's**

| Milestone / Authors | Year | Mathematical Basis | Required Inputs / Calibration | Key Scientific Contribution |
|---|---|---|---|---|
| **Shen & Glinka**[21,22] | 1991 | Universal fractional power series: $\left(1-\frac{x}{a}\right)^{21/}, \left(1-\frac{x}{a}\right)^{1},$ $\left(1-\frac{x}{a}\right)^{23/2}$ | Two reference $K_I$ solutions + Fett's crack-mouth curvature condition $\left[\frac{\partial^2 m}{\partial a^2}\right]_{x=0} = 0$ | Improved approximation accuracy for steep, non-linear stress gradients (e.g., notches, residual stresses). |
| **Wu & Carlsson**[23] | 1991 | Bivariate integer power series: $M_i = f\left(a/c; a/t\right)$ | Comprehensive 3D FEA stress intensity factor and displacement databases | Extended integer-power weight functions to 3D semi-elliptical surface cracks in plates and shells. |
| **Wang & Lambert**[24] | 1995 | Bivariate fractional power series for deepest point (A) & surface point (B) | Reference SIF solutions (Newman–Raju) + FEA datasets | Established the industrial standard for 3D semi-elliptical surface cracks, adopted by BS 7910, API 579, and R6. |

## 7. Results: thermally strengthened glass.

The application of the weight functions method to glass[26,27], and specifically to strengthened glass, is quite natural considering the mechanical behavior of glass dominated by the surface flaws/cracks population. In the present study, we consider a 4mm glass plate with a critical stress intensity factor of $K_{IC}$=0.75 MPa√m and $K_{ITh}$=0.25 MPa√m. We consider 2D edge cracks. The first case study is annealed glass that is glass where $\sigma_{res}(x)=0$. When this glass is submitted to a uniform tensile stress $SM=50\ MPa$, the $K_I(a)$ is from equation (1):

$$K_I UN(a) = Y \cdot SM \cdot \sqrt{a}\,. \tag{25}$$

Considering an externally applied pure bending load with maximum bending stress of $SM=50\ MPa$, the non-uniform stress field is represented by equation (7):

$$\sigma_{appl}(x) = SM\left(1 - 2\frac{x}{d}\right), \tag{26}$$

and the $K_I(a)$ results:

$$K_I\left(a\right) = \int_0^a \sigma_{appl}(x) \cdot m(x,a)dx\,. \tag{27}$$

In Figure 12 the uniform stress intensity factor $K_IUN(a)$ is reported compared with the $K_I(a)$ resulting from the non-uniform bending stress. It can be noted that the two curves are practically coincident up to about a crack of 50 micron depth, while after this point, the two curves remain quite similar. This is due to the point that the change in applied stress (Equation (26)) is not significant for shallow surface cracks below 100 μm depth. It can also be noted that $K_I(a)$ is a positive growing function of $a$. The $K_I(a)$ curve intersects the $K_{Ith}$ curve at a value of $a$ below 10μm. This means that at this point and in an environment with moisture, the surface crack activates the subcritical crack growth mechanism. This point represents an instability condition for the crack depth and shall be avoided to guarantee a residual service life of the glass article. The value of 10 microns depth for a surface flaw is very critical; in normal handling during manufacturing processes, surface flaws can reach about 20 microns, and in service life, depending on application conditions, surface damages may easily exceed this shallow level.

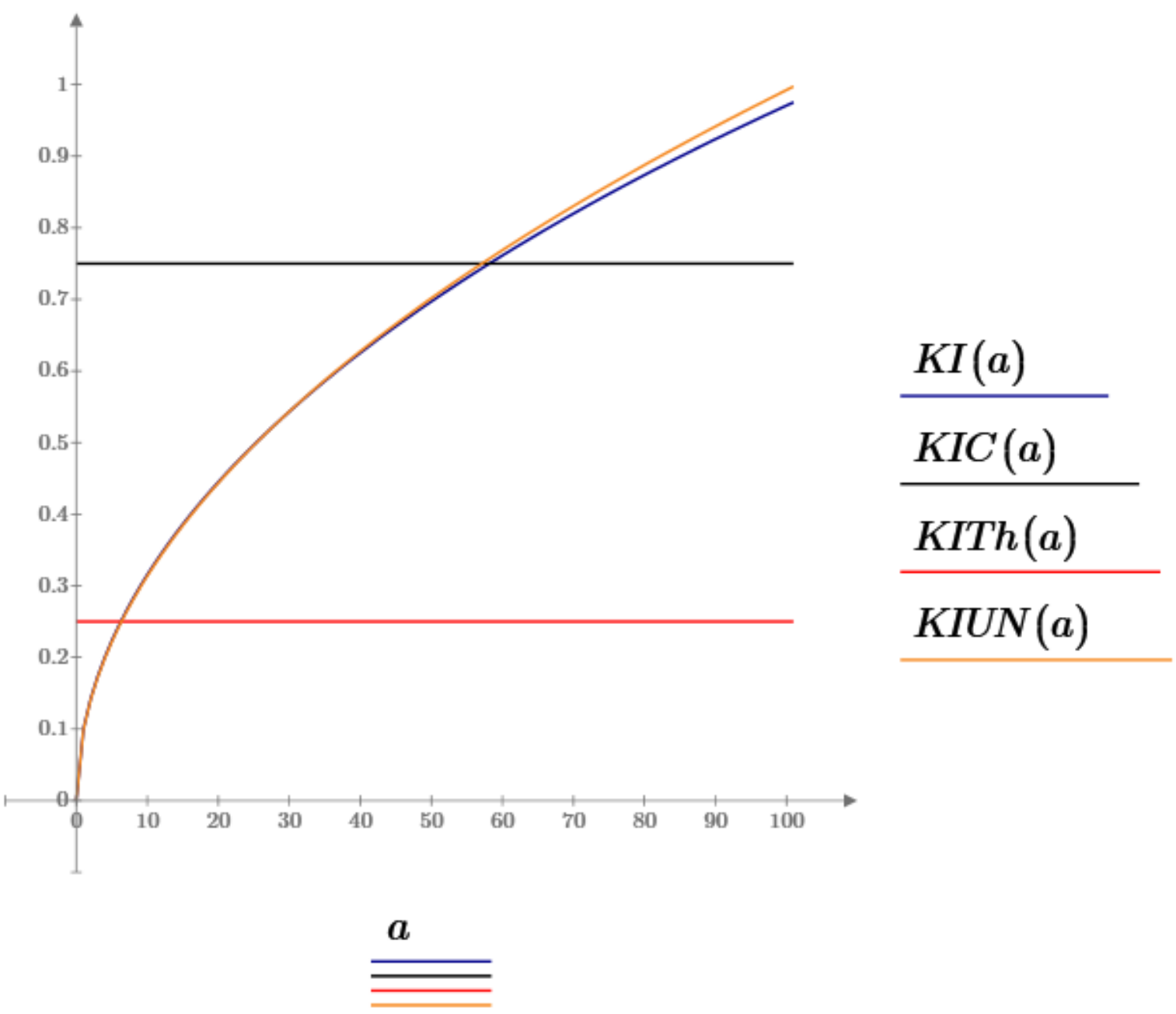

Figure 12 – Annealed glass under a uniform tensile stress KIUN(a) of 50 MPa and a non-uniform bending stress field KI(a) with a surface tensile stress value of 50 MPa. Curves are compared with the threshold SGG value and the critical value $K_{IC}$.

To overcome potential strength issues of annealed glass, glass articles can be thermally strengthened according to what is indicated in section 5. Introducing a residual stress represented by equation (10) and figure 8 (*Sc=90 MPa, CT=45 MPa*) the calculated *KITSG (a)* curve is reported in Figure 13. The introduction of residual stress significantly changes the $K_I(a)$ curve, identifying 3 important areas:

A) value of $K_I(a_{min})$ where $K_I(a)$ is at a minimum of $K_I\left(\frac{dK_I(a_{min})}{da}=0\right) < K_I(a) < 0$

B) From $a_{min}$ to the value of $a$ where $K_I(a)=0$

C) $0 < K_I(a) < K_{Ith}$

The presence of a negative value of $K_I(a)$ and of a part of the curve with a negative derivative indicates an area of great stability. Glass is also quite stable for the region identified for damage below a crack depth where $K_I(a)<0$. The critical limit is reached when the $K_I(a)$ value approaches the $K_{ITh}$ point.

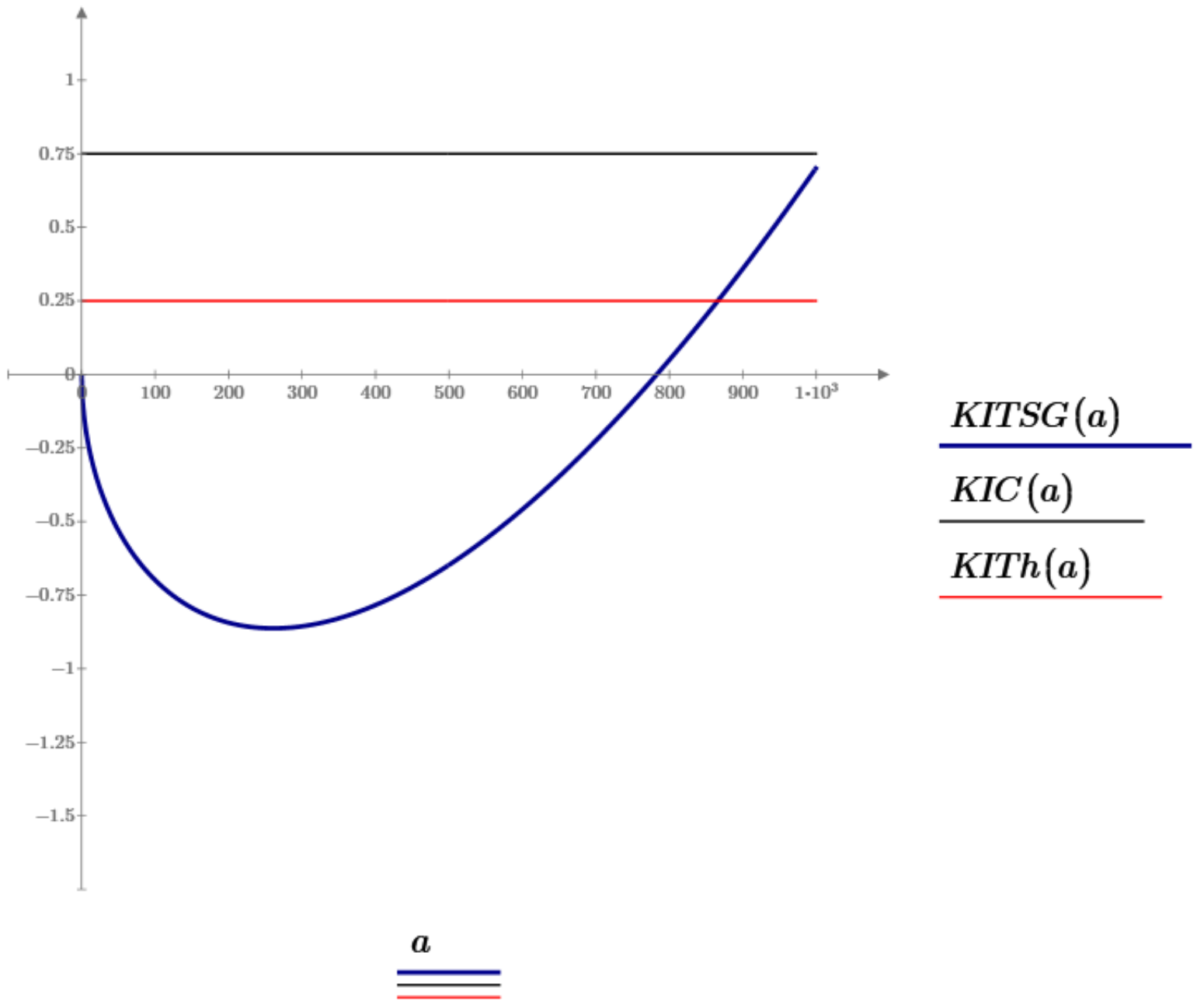

Figure 13 – Stress intensity factor diagram $K_I(a)$ for a thermally strengthened glass (residual stress profile according to Figure 8) under a non-uniform bending stress field with a surface tensile stress value of 50 MPa. Curves are compared with the threshold SGG value and the critical value $K_{IC}$.

## 8. Results: chemically strengthened glass

The application of the weight functions method to chemically strengthened glass can be traced to some literature references: Abrams and Green[26] and Egboiyi and others[27]. In the present study, two cases are considered as indicated in section 5: shallow depth of compression on soda lime silicate glass (SLS) and deep compression layer on Sodium aluminosilicate glass (SAS). The resulting $K_I(a)$ curves for the two investigated cases are reported in Figures 14 and 15.

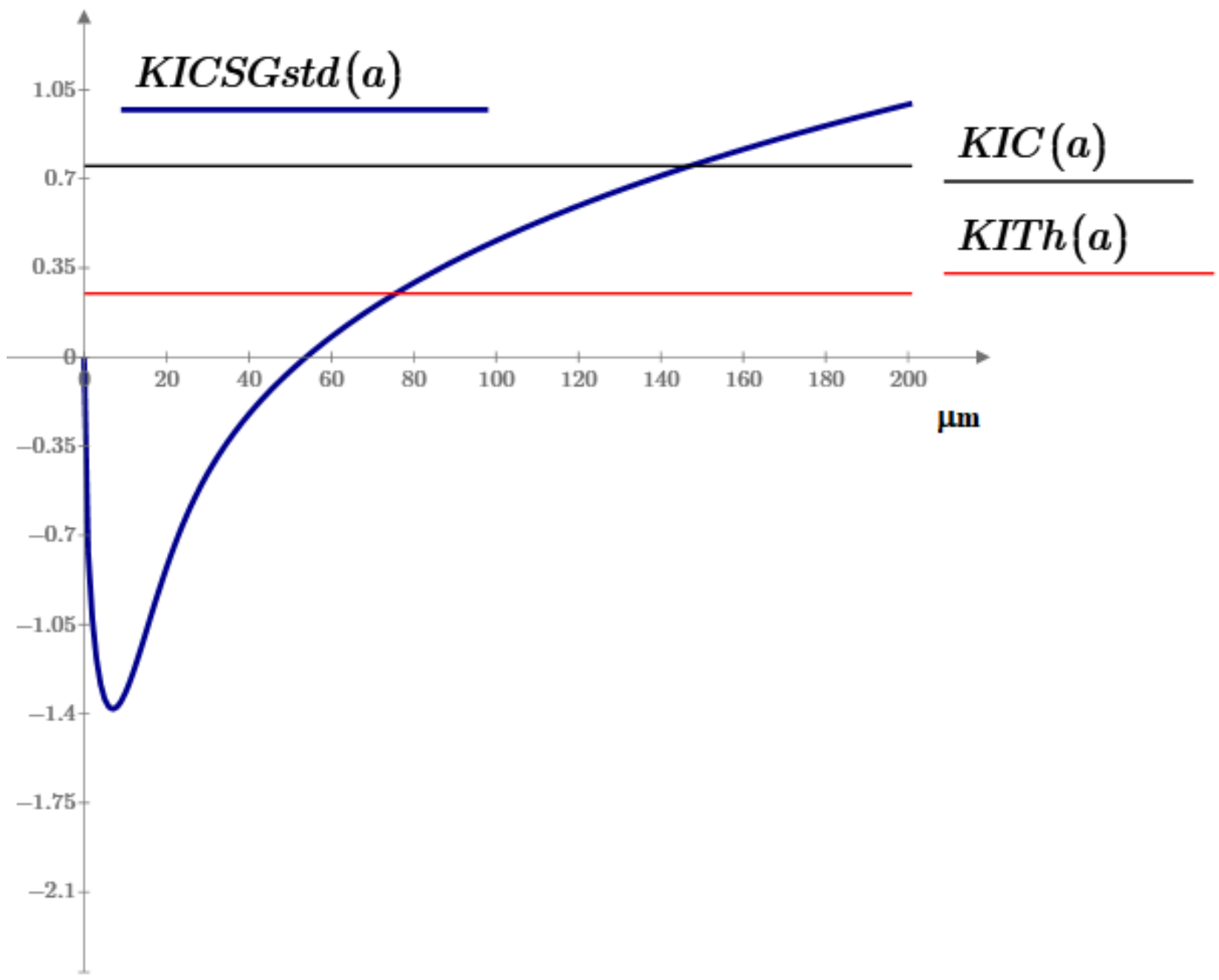

Figure 14 – Stress intensity factor diagram $K_I(a)$ for a chemically strengthened SLS glass (residual stress profile according to Figure 10) under a non-uniform bending stress field with a surface tensile stress value of 50 MPa. Curves are compared with the threshold SGG value and the critical value $K_{IC}$.

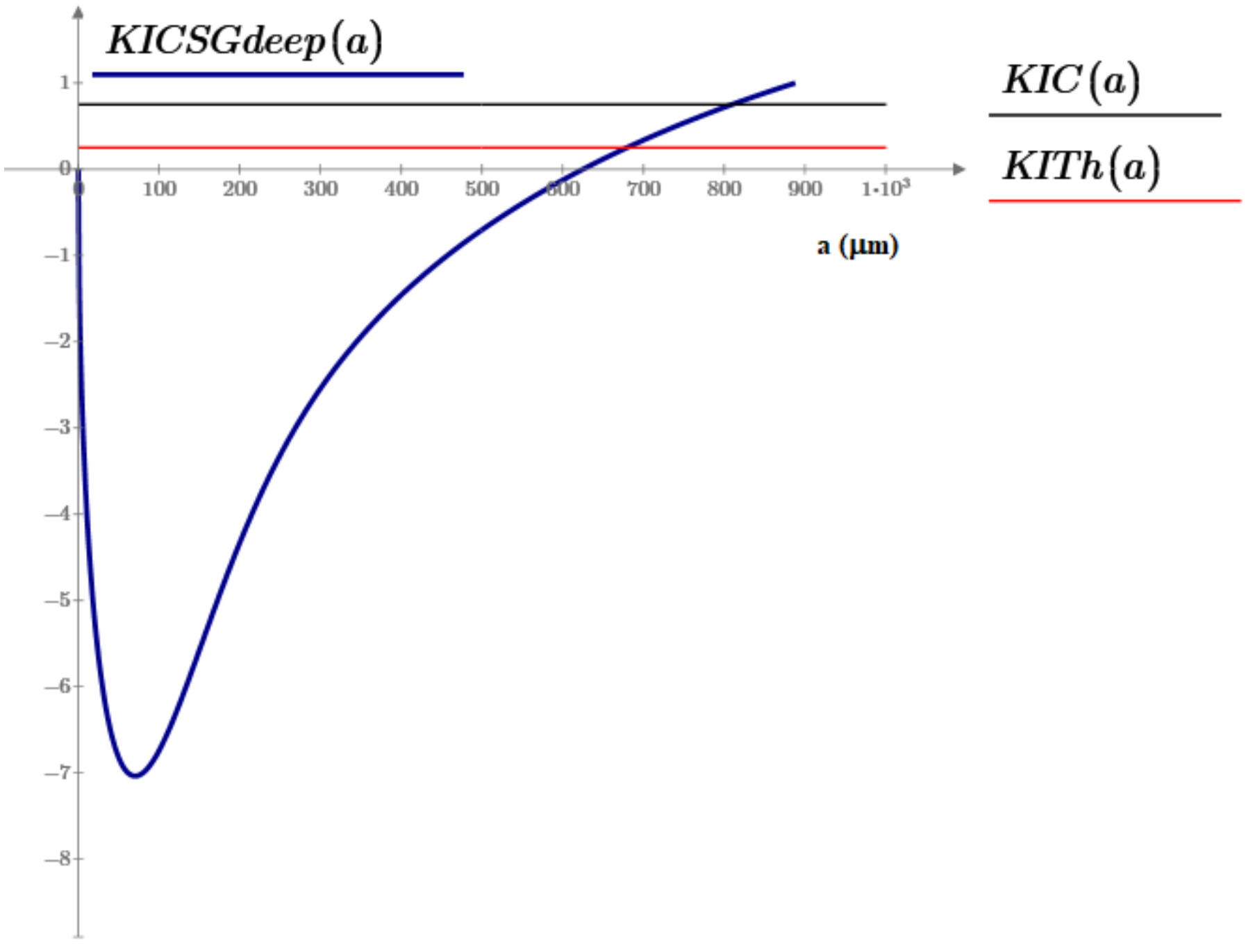

Figure 14 – Stress intensity factor diagram $K_I(a)$ for a chemically strengthened SAS glass (residual stress profile according to Figure 10) under a non-uniform bending stress field with a surface tensile stress value of 50 MPa. Curves are compared with the threshold SGG value and the critical value $K_{IC}$.

Comparing Figure 14 to Figure 15 it can be observed that the significant increase of compression layer depth Cd increases stability conditions by about an order of magnitude. The $K_I(a)$ curve intersects the zero axis between 40 and 60 µm for standard CSG, while the same intersection is found around 600 µm for the deep CSG. This evidence has great relevance for structural glass applications where severe surface actions (indentations, impacts, scratches) may lead to deep surface crack formation.

## 9. Conclusion

The stability conditions in the KI(a) curves for thermally strengthened glass and chemically strengthened glass have been identified:

A) Full stability condition when the derivative of $K_I(a)$ ($dK_I/da \leq 0$) is negative

B) Conservative stability for values of a where $K_I$ remains negative - $K_I(a) \leq 0$

C) Limit stability condition for positive values of $K_I(a)$ up to the SCG onset ($K_I(a)=K_{Ith}$).

The limit condition represented by the C) line shall be considered very carefully. As already pointed out, when $K_I(a)$ approaches $K_{Ith}$ , the surface crack starts to be activated by the subcritical growth mechanism and, at this point, it becomes critical to have control of the crack growth dynamical evolution.

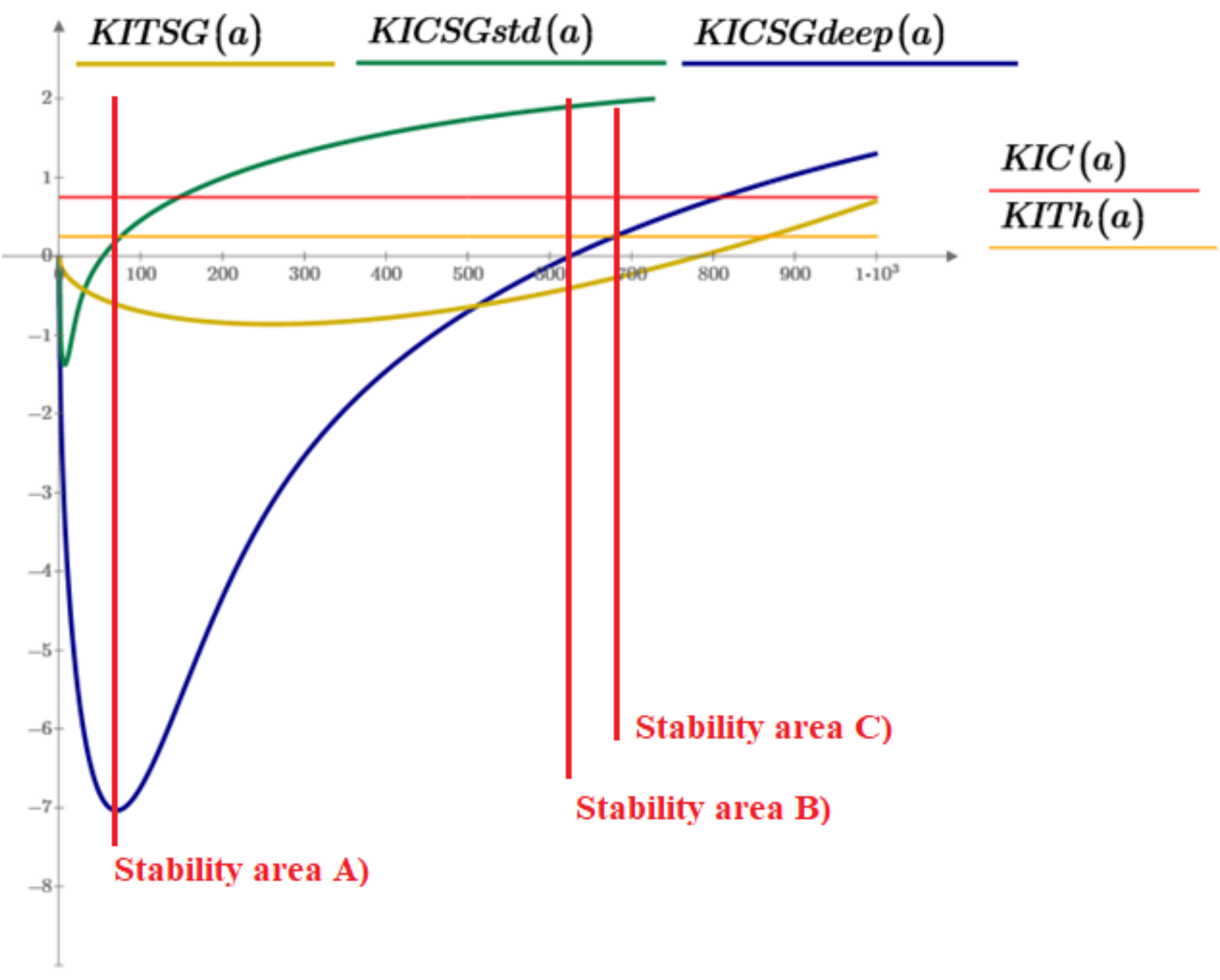

Figure 15 – Comparison of stress intensity factor diagrams $K_I(a)$ for thermally strengthened glass (TSG) and the two types of chemically strengthened glass (CSG_DEEP for SAS glass and CSG_STD for SLS glass) under a non-uniform bending stress field with a surface tensile stress value of 50 MPa. Curves are compared with the threshold SGG value and the critical value $K_{IC}$. The stability areas according to Table VI are indicated only for the CSG_DEEP glass.

The three different $K_I(a)$ diagrams calculated for the three studied glass types are compared in Figure 15 where the stability areas are indicated for the CSG_DEEP type. Values of crack depth that identify the stability areas are reported in Table VI for the three different glass types. The application of glass in service life scenarios where surface damage effects may occur, suggests carefully considering the potential extension of such defects and consequently choosing the most appropriate strengthening process to ensure the structural integrity of the glass article during its service life. This study is limited to 2D continuous edge cracks. An extension to 3D semi-elliptical surface cracks is recommended to evaluate how crack aspect ratios a/c and a/d may have an impact to the mechanical behaviour of the glass article in different strengthening and service load conditions.

**TABLE VI – Stability condition areas and stability values (in terms of depth of *a*) for the three types of glass considered in this study.**

| Stability Condition | A) $dK_I/da \leq 0$ | B) $K_I(a) \leq 0$ | C) $K_I(a) < K_{Ith}$ |
|---|---|---|---|
| Strengthening | | | |
| TSG | 260 μm | 782 μm | 866 μm |
| CSG-STD | 7 μm | 53 μm | 75 μm |
| CSG-Deep | 70 μm | 620 μm | 680 μm |

**APPENDIX 1**

**Proof of some approximations derived from the residual stress equation for ion-exchanged glasses.**

The fundamental stress profile equation introduced in the text:

$$\sigma_{EB}(x,t) = -\frac{ScV}{C_s}\left[\left(c(x,t)-\bar{c}(t)\right)-\int_0^t \frac{\partial R(t-\theta)}{\partial\theta}\left(c(x,\theta)-\bar{c}(\theta)\right)d\theta\right], \tag{A1}$$

$$R(t) = \exp\left[-\left(\frac{t}{\tau}\right)^b\right], \tag{A2}$$

was derived[16] from the Sane-Cooper equation[15]

$$\sigma_{EB}(x,t) = -\frac{ScV}{C_s}\left[\int_0^t R(t,\theta)\cdot\frac{\partial}{\partial\theta}\left(c(x,\theta)-\bar{c}(\theta)\right)d\theta\right] \quad , \tag{A3}$$

using the integration by parts formula. This means that equations (A1) and (A3) are equivalent. The evolution of surface compression over time can be directly evaluated from (A1):

$$Sc(t) = \sigma_{EB}(0,t) = -\frac{ScV}{C_s}\left[\left(c(0,t)-\bar{c}(t)\right)-\int_0^t \frac{\partial R(t-\theta)}{\partial\theta}\left(c(0,\theta)-\bar{c}(\theta)\right)d\theta\right]. \tag{A4}$$

Let's define *$c(0,t)=C_s$,* which is the concentration at the glass surface that is the boundary condition of constant surface concentration over time. The assumption we consider is:

$$C_s - \bar{c}(t) \approx C_s \ , \tag{A5}$$

that is, the concentration at the surface is much higher than the average concentration. This approximation is reasonable when considering that the ion-exchanged layer is usually much thinner than the overall glass article thickness. Considering approximation (A5) in equation (A4) results:

$$Sc(t) = -\frac{ScV}{C_s}\left[C_s - \int_0^t \frac{\partial R(t-\theta)}{\partial\theta}C_s\right] = -\frac{ScV}{C_s}\left[C_s - C_s\int_0^t \frac{\partial R(t-\theta)}{\partial\theta}d\theta\right], \tag{A6}$$

Because R(0)=1, considering equation (A6) results:

$$Sc(t) = -ScV\cdot R(t) \tag{A7}$$

Equation (A7) proves equation (18).

The second approximation is based on the possibility of decoupling the convolution integral in equations (A1) and (A3) by the following position:

$$R(t-\theta) \approx R(t), \tag{A8}$$

this more drastic approximation has the physical meaning of losing the memory in the convolution integral. This approximation considers a relaxation process where the relaxation time $\tau$ is much slower than the observation time $\tau << t$.. Introducing equation (A8) directly into the Sane-Cooper equation (A3) results:

$$\sigma_{EB}(x,t) \approx -\frac{ScV}{C_s} \cdot R(t)\left[\int_0^t \frac{\partial}{\partial\theta}\left(c(x,\theta)-\bar{c}(\theta)\right)d\theta\right], \tag{A9}$$

so that:

$$\sigma_{EB}(x,t) \approx -\frac{ScV}{C_s} \cdot R(t) \cdot \left[c(c,t-\bar{c}(t)\right] = -\frac{Sc(t)}{C_s}\left[c(c,t-\bar{c}(t)\right]. \tag{A10}$$

This proves equation (19).